\documentclass[11pt]{article}

\usepackage[font=small,labelfont=bf]{caption}

\usepackage{enumitem}
\usepackage{graphicx}
\usepackage{xcolor}
\usepackage{blkarray}
\usepackage{amsmath,amssymb}
\usepackage{bbm}
\usepackage{mathtools}
\usepackage{pdflscape}
\usepackage{float}
\usepackage{rotating}
\usepackage{titlesec}

\usepackage[
citestyle=authoryear,
style=authoryear,
uniquename=false,
sorting=ynt,
maxbibnames=8
]{biblatex}

\titleformat{\paragraph}
{\normalfont\normalsize\bfseries}{\theparagraph}{1em}{}
\titlespacing*{\paragraph}
{0pt}{3.25ex plus 1ex minus .2ex}{1.5ex plus .2ex}

\DeclareMathOperator{\EX}{\mathbbm{E}}% expected value

\begin{document}
\setlength{\abovedisplayskip}{4pt}
\setlength{\belowdisplayskip}{10pt}
\setlength{\abovedisplayshortskip}{4pt}
\setlength{\belowdisplayshortskip}{10pt}

\title{Hypnozoite dynamics for \textit{Plasmodium vivax} malaria: the epidemiological effects of radical cure}

\author{Somya Mehra$^1$ \and Eva Stadler$^2$ \and David Khoury$^2$ \and James M. McCaw$^{1,3,4}$ \and Jennifer A. Flegg$^1$}

\date{%
    $^1$School of Mathematics and Statistics, The University of Melbourne, Parkville,  Australia\\%
    $^2$Kirby Institute, University of New South Wales, Kensington, Australia\\
    $^3$Centre for Epidemiology and Biostatistics, Melbourne School of Population and Global Health, The University of Melbourne, Parkville, Australia\\
    $^4$Peter Doherty Institute for Infection and Immunity, The Royal Melbourne Hospital and The University of Melbourne, Parkville, Australia\\[2ex]%
}

\maketitle

\section*{Abstract}
Malaria is a mosquito-borne disease with a devastating global impact. \textit{Plasmodium vivax} is a major cause of human malaria beyond sub-Saharan Africa. Relapsing infections, driven by a reservoir of liver-stage parasites known as hypnozoites, present unique challenges for the control of \textit{P. vivax} malaria. Following indeterminate dormancy periods, hypnozoites may activate to trigger relapses. Clearance of the hypnozoite reservoir through drug treatment (radical cure) has been proposed as a potential tool for the elimination of \textit{P. vivax} malaria. Here, we introduce a stochastic, within-host model to jointly characterise hypnozoite and infection dynamics for an individual in a general transmission setting, allowing for radical cure. We begin by extending an existing activation-clearance model for a single hypnozoite, adapted to both short- and long-latency strains, to include drug treatment. We then embed this activation-clearance model in an epidemiological framework accounting for repeated mosquito inoculation and the administration of radical cure. By constructing an open network of infinite server queues, we derive analytic expressions for several quantities of epidemiological significance, including the size of the hypnozoite reservoir; the relative contribution of relapses to the infection burden; the distribution of multiple infections; the cumulative number of recurrences over time, and the time to first recurrence following drug treatment. By deriving, rather than assuming parameteric forms, we characterise the transient dynamics of the hypnozoite reservoir following radical cure more accurately than previous approaches. To yield population-level insights, our analytic within-host distributions can be embedded in multiscale models. Our work thus contributes to the epidemiological understanding of the effects of radical cure on \textit{P. vivax} malaria. 

\section{Introduction}
Malaria remains a significant cause of morbidity and mortality, with an estimated 229 million cases and 409,000 deaths in 2019 alone \parencite{world2020world}. \textit{Plasmodium falciparum} and \textit{Plasmodium vivax}, which are transmitted to humans through the bites of infected \textit{Anopheles} mosquitoes, are the primary contributors to the global malaria burden. While the prevalence of \textit{P. falciparum} remains higher than that for \textit{P. vivax} globally, the relative burden of \textit{P. vivax} has increased in various co-endemic settings as malaria elimination efforts have intensified \parencite{price2020plasmodium}.\\ 

The control and elimination of \textit{P. vivax} malaria is complicated by various biological characteristics of both parasite and vector \parencite{world2015control, howes2016global, olliaro2016implications, price2020plasmodium}. Relapsing infections, driven by a reservoir of latent liver-stage parasites known as hypnozoites, are perhaps the key distinguishing feature of \textit{P. vivax}. For vivax malaria, each infective bite can trigger a primary (blood-stage) infection, in addition to establishing a variable hypnozoite inoculum \parencite{white2012relapse}. Hypnozoites remain inactive and undetectable in the liver for indeterminate periods, with long-latency phenotypes typically observed in temperate regions, and short-latency phenotypes generally observed in tropical regions \parencite{white2012relapse, battle2014geographical}. Each hypnozoite activation event, however, has the potential to trigger a new blood-stage infection, called a relapse. The hypnozoite reservoir can thus re-establish transmission within a community, even after the elimination of all active infections \parencite{shanks2012control}.\\

Although the clearance of the hypnozoite reservoir is critical to elimination efforts, the majority of antimalarial drugs exclusively target the blood-stages of infection. Only a small class of drugs have hypnozonticidal activity; such treatments are collectively referred to as radical cure because of their ability to eliminate both active and latent parasites \parencite{wells2010targeting}. Radical cure has been proposed as a potential tool for \textit{P. vivax} malaria elimination \parencite{shanks2012control}. However, the widespread adoption of radical cure treatments has been curtailed, largely because of the risk of haemolysis in G6PD deficient individuals \parencite{wells2010targeting, world2015control}. Here, we develop a mathematical model of the dynamics of the hypnozoite reservoir within a single host in a general transmission setting to explore the epidemiological consequences of radical cure.\\

Previous simulation models of within-host \textit{P. vivax} dynamics have examined the complexities of hypnozoite activation and blood-stage infection, but have not considered the accrual of the hypnozoite reservoir in endemic settings \parencite{kerlin2015simulation}. Transmission models accounting for individuals carrying hypnozoites have been developed, but have generally compartmentalised individuals carrying at least one hypnozoite, without explicitly modelling the size of the hypnozoite reservoir \parencite{ishikawa2003mathematical, aguas2012modeling, roy2013potential, chamchod2013modeling, robinson2015strategies, white2016variation}. Various distributional forms for the time-to-relapse, including exponential distributions \parencite{aguas2012modeling, chamchod2013modeling,robinson2015strategies, white2016variation}, log-normal distributions \parencite{ishikawa2003mathematical}, gamma distributions \parencite{roy2013potential} and mixture distributions \parencite{lover2014distribution, taylor2019resolving}, have been assumed without accounting for the dependency between the size of the hypnozoite reservoir and the risk of relapse. Transmission models accounting for the accrual of the hypnozoite reservoir over successive mosquito bites, in addition to immunity, prophylaxis and clinical symptoms, have been proposed by  \textcite{white2018mathematical}; yet, in assuming that each batch of hypnozoites (established by the same mosquito bite) gives rise to relapses at the same constant rate, these models do not account for variability in parasite inocula across bites, which can modulate the risk of relapse \parencite{white2012relapse}. By embedding a within-host model of hypnozoite activation in a population-level transmission model accounting for variability in hypnozoite inocula, \textcite{white2014modelling} have obtained distributions for the prevalence of vivax malaria and the size of the hypnozoite reservoir under a range of control interventions, including radical cure. However, distributions of multiple infections; the relative contributions of primary infections to the infection burden; and the cumulative number of relapses over time have not been examined in this framework, which has moreover been restricted to short-latency strains \parencite{white2014modelling}.\\ 

In this paper, we develop a within-host model to jointly characterise the accrual of the hypnozoite reservoir and the infection burden over time, whilst accounting for drug treatment, for both short- and long-latency strains. In Section \ref{sec::single_hyp}, we extend an existing activation-clearance model for a single hypnozoite \parencite{white2014modelling, mehra2020activation} to consider treatment with blood-stage (schizontocidal) drugs and radical cure. To characterise the dynamics of the hypnozoite reservoir and blood-stage infections in an endemic setting, we then embed this activation-clearance model in an epidemiological framework in Section \ref{sec::inf_server_queue}, extending our previous work \parencite{mehra2021antibody} to account for drug treatment. By constructing an open network of infinite server queues, we derive a joint probability generating function (PGF) for the number of hypnozoites in each state of the model, in addition to the number of cleared and ongoing recurrences, for an individual in a general transmission setting. In Section \ref{sec::epi_q}, we derive analytic expressions for quantities of epidemiological significance, including the size of the hypnozoite reservoir; the risk of primary infections and relapses over time; the incidence of multiple infections; the time to first recurrence following drug treatment and the cumulative number of recurrences in a given interval. To capture the epidemiological effects of radical cure, we compare hypnozoite and infection dynamics following treatment with radical cure against a scenario with no treatment or blood-stage treatment only, with illustrative results provided and discussed in Section \ref{sec::illustrative_results} and concluding remarks in Section \ref{sec::discussion}.

\section{Relapse-Clearance Dynamics for a Single Hypnozoite} \label{sec::single_hyp}

\subsection{Baseline Scenario}
\label{sec::baseline_hyp}
\begin{figure}
    \centering
    \includegraphics[width=\textwidth]{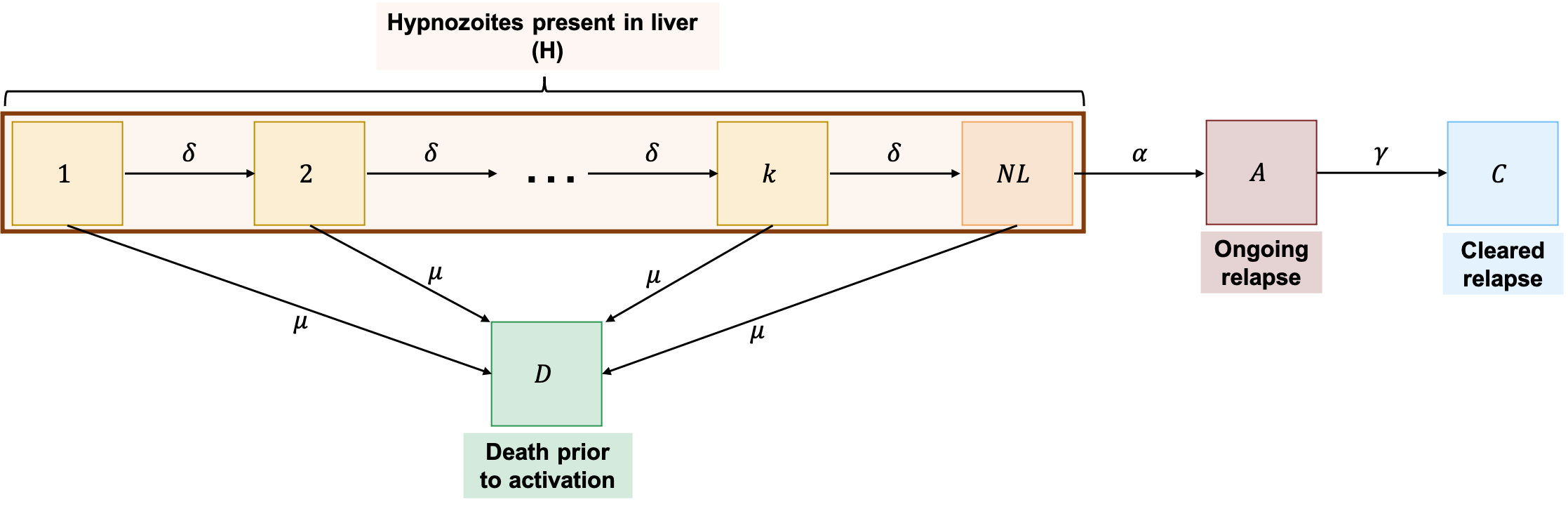}
    \caption{Schematic for relapse-clearance model of a single hypnozoite under a baseline scenario, in the absence of drug treatment or external triggers of hypnozoite activation. States $1, \dots, k$ denote latency compartments; $NL$ denotes a non-latent hypnozoite; $D$ denotes a hypnozoite that has died prior to activation; $A$ denotes an ongoing relapse triggered by hypnozoite activation and $C$ denotes a relapse that has been cleared from the bloodstream. State $H$ collectively refers to hypnozoites that are present in the liver, that is, both non-latent ($NL$) and latent ($1, \dots, k$) hypnozoites. Setting $k=0$ captures the dynamics of short-latency (tropical) strains, while $k>0$ applies to long-latency (temperate) strains \parencite{white2014modelling}.}
    \label{fig:single_hyp}
\end{figure}

We begin by developing a model of relapse-clearance dynamics for a single hypnozoite in a baseline scenario, neither accounting for drug treatment nor external triggers of hypnozoite activation. Similarly to \textcite{white2014modelling}, we assume that each hypnozoite undergoes a dormancy phase, during which it can die, but not activate. We model this dormancy phase as a series of $k \geq 0$ compartments, with transition rate $\delta$ between compartments.  Hypnozoites that have emerged from dormancy (hereafter referred to as non-latent hypnozoites) are assumed to activate at some constant rate $\alpha$. We also assume that all hypnozoites in the liver (that is, both dormant and non-latent hypnozoites) are subject to death at constant rate $\mu$, potentially due to the death of  the host hepatocyte. This activation-clearance model was introduced in \textcite{white2014modelling} and discussed in detail in \textcite{mehra2020activation}, along with analytic solutions to the state probabilities. Here, we further assume that hypnozoite activation immediately triggers a blood-stage infection (relapse) that is cleared at rate $\gamma$ (that is, exponentially-distributed with expected duration $1/\gamma$). Each hypnozoite therefore has two possible end states: death prior to activation, or clearance following blood-stage infection. A schematic of this model structure is shown in Figure \ref{fig:single_hyp}.\\

The case $k>0$ captures hypnozoite dynamics for long-latency (temperate) strains of \textit{P. vivax}. By setting $k=0$, that is, accounting for a scenario where each hypnozoite may activate immediately after it is established in the liver, we recover a model for short-latency (tropical) strains \parencite{white2014modelling}.\\

Suppose that a single hypnozoite is established in a host hepatocyte at time zero. Denote the state of the hypnozoite at time $t$ by $X(t)$. Then $X(t)$ has probability mass function (PMF)
\begin{align*}
    \mathbf{p}(t) = (p_1(t), \dots, p_k(t), p_{NL}(t), p_A(t), p_C(t), p_D(t))
\end{align*}
where the state probabilities are defined to be
\begin{itemize}
    \item $p_m(t)$ that the hypnozoite is present in latency compartment $m \in [1, k]$ at time $t$;
    \item $p_{NL}(t)$ that a hypnozoite is non-latent, that is, present in the liver and may activate, at time $t$ (state $NL$);
    \item $p_A(t)$ that the hypnozoite has activated and triggered a relapse that is ongoing at time $t$ (state $A$);
    \item $p_C(t)$ that the hypnozoite has activated to cause a relapse that has been cleared by time $t$ (state $C$);
    \item $p_D(t)$ that the hypnozoite has died prior to activating by time $t$ (state $D$).
\end{itemize}

For notational convenience, we define
\begin{align*}
    p_H(t) = \sum^{k}_{m=1} p_m(t) + p_{NL}(t)
\end{align*}
to be the probability that a hypnozoite is present in the liver, that is, state $H$, at time $t$.\\

Based on the model schematic in Figure \ref{fig:single_hyp} and the Kolmogrov forward differential equations, it follows that
\begin{align}
    \frac{dp_1}{dt} &= -(\delta + \mu) p_1(t) \label{deq_L1} \\
    \frac{dp_m}{dt} &= -(\delta + \mu) p_m(t) + \delta p_{m-1}(t), \, m \in [2, k] \label{deq_Li} \\
    \frac{dp_{NL}}{dt} &= -(\alpha + \mu) p_{NL}(t) + \delta p_k(t) \label{deq_NL} \\
    \frac{dp_{A}}{dt} &= - \gamma p_A(t) + \alpha p_{NL}(t) \label{deq_A} \\
    \frac{dp_C}{dt} &= \gamma p_A(t)\\
    \frac{dp_D}{dt} &= \mu \sum^{k}_{i=1} p_i(t) + \mu p_{NL}(t) = \mu p_H(t) \label{deq_D},
\end{align}
with the initial condition
\begin{small}
\begin{align}
    \mathbf{p}(0) = \begin{cases}
    (p_1(0), p_2(0), \dots, p_k(0), p_{NL}(0), p_A(0), p_C(0), p_D(0)) = (1, 0, \dots, 0, 0, 0, 0, 0) & \text{ if } k>0\\
    (p_{NL}(0), p_A(0), p_C(0), p_D(0)) = (1, 0, 0, 0) & \text{ if } k=0
    \end{cases}\label{ic}
\end{align}
\end{small}

Integrating by parts, we can solve the system in Equations (\ref{deq_L1}) to (\ref{deq_D}), subject to initial condition (\ref{ic}) to yield
\begin{align}
	p_{m}(t) =& \frac{(\delta t)^{m-1}}{(m-1)!}e^{-(\mu+\delta)t} \text{ for } m \in [1, k] \label{l_eq}\\
	p_{NL}(t) =& \frac{\delta^{k}}{(\delta - \alpha)^{k}} \Bigg[ e^{-(\mu + \alpha)t} - e^{-(\mu + \delta)t} \sum^{k-1}_{j=0} \frac{t^j}{ j!} (\delta - \alpha)^{j} \Bigg] \label{nl_eq} \\
    p_A(t) =& \frac{\alpha \delta^{k}}{(\delta - \alpha)^{k}} \Bigg[\frac{e^{-(\mu + \delta)t}}{\mu - \gamma + \delta} \Bigg\{ \sum^{k-1}_{j=0} \Big( \frac{\delta - \alpha}{\mu - \gamma + \delta} \Big)^j \sum^{j}_{i=0} \frac{t^i}{i!} (\mu -\gamma + \delta)^{i} \Bigg\} - \frac{e^{-(\mu + \alpha)t}}{\mu - \gamma + \alpha} \Bigg] + \notag \\
    & \frac{\alpha}{\alpha + \mu - \gamma} \Big( \frac{\delta}{\delta + \mu - \gamma} \Big)^{k} e^{-\gamma t} \label{a_eq} \\
    p_C(t) = & \frac{\alpha}{\alpha + \mu} \Big( \frac{\delta}{\mu + \delta} \Big)^k -\frac{\alpha}{\alpha + \mu - \gamma} \Big( \frac{\delta}{\delta + \mu - \gamma} \Big)^{k} e^{-\gamma t} + \frac{\gamma \alpha \delta^{k}}{(\delta - \alpha)^{k}} \frac{e^{-(\mu + \alpha)t}}{(\mu + \alpha)(\mu - \gamma + \alpha)} - \notag \\
    & \frac{\gamma \alpha \delta^{k}}{(\delta - \alpha)^{k}} \frac{e^{-(\mu + \delta)t}}{(\mu - \gamma + \delta)(\mu + \gamma)} \Bigg\{ \sum^{k-1}_{j=0} \Big( \frac{\delta - \alpha}{\mu - \gamma + \delta} \Big)^j \sum^{j}_{i=0} \Big( \frac{\mu - \gamma + \delta}{\mu + \delta} \Big)^{i} \sum^i_{q=0} \frac{t^q}{q!} (\mu + \delta)^q \Bigg\} \label{c_eq}\\
    p_D(t) =& 1 - \sum^k_{m=1} p_m(t) - p_{NL}(t) - p_A(t) - p_C(t), \label{d_eq}
\end{align}
where we have used standard integral number 2.321.2 in \textcite{jeffrey2007table}. For a physical interpretation of the activation-clearance system for long-latency strains, which accounts for hypnozoite death, dormancy and activation, but does not account for the clearance of relapses (that is, does not distinguish states $A$ and $C$), see \textcite{mehra2020activation}.\\

In the case $k=0$, corresponding to short-latency (tropical) strains, whereby all hypnozoites in the liver may activate, Equations (\ref{nl_eq}) to (\ref{d_eq}) simplify to
\begin{align}
    p_H(t) &= p_{NL}(t) = e^{-(\alpha + \mu)t}\\
    p_A(t) &= \frac{\alpha}{(\alpha+\mu) - \gamma} \big( e^{-\gamma t} - e^{-(\alpha + \mu)t} \big)\\
    p_C(t) &= \frac{\alpha}{\alpha + \mu} \big( 1 - e^{-(\alpha + \mu)t} \big) - \frac{\alpha}{(\alpha+\mu) - \gamma} \big( e^{-\gamma t} - e^{-(\alpha + \mu)t} \big) \\
    p_D(t) &= \frac{\mu}{\alpha + \mu} \big( 1 - e^{-(\alpha + \mu)t} \big).
\end{align}

\subsection{Drug Treatment}
We now extend the baseline model introduced in Section \ref{sec::baseline_hyp} to account for drug treatment. We make the simplifying approximation that drug treatment has an instantaneous effect; while antimalarial drug half-lives vary broadly, from approximately 40 minutes for artesunate \parencite{morris2011review}, to 6 hours for primaquine (radical cure) \parencite{white1992antimalarial} and 30-60 days for chloroquine \parencite{white1992antimalarial}, we are primarily concerned with hypnozoite dynamics over a time frame of years, and therefore this assumption of instantaneous action is appropriate. Upon administration of drug treatment, we thus assume that each hypnozoite in the liver (that is, states $1, \dots, k, NL$, collectively referred to as state $H$) dies instantaneously (that is, transitions to state $D$) with probability $p_\text{rad}$; while any ongoing blood-stage infections (that is, hypnozoites in state $A$) are instantaneously cleared (that is, transition to state $C$) with probability $p_\text{blood}$. The case $p_\text{blood}>0$, $p_\text{rad}=0$ corresponds to blood-stage treatment only, while $p_\text{blood}>0$, $p_\text{rad}>0$ corresponds to hypnozonticidal treatment (radical cure). Any hypnozoites that survive radical cure (that is, remain in state $H$) or persisting blood-stage infections (state $A$) are then subject to the same activation-clearance dynamics described in Section \ref{sec::baseline_hyp}.\\

Consider a single hypnozoite inoculated at time $t=0$. Suppose that drug treatment is administered successively at times $s_1$, $s_2$, \dots, $s_n$. We denote the state of the hypnozoite at time $t$ $X^r(t, s_1, \dots, s_n) \in \{ 1, \dots, k, NL, A, C, D \}$ with corresponding PMF $\mathbf{p^r}(t, s_1, \dots, s_n)$. The governing equations for the state probabilities are given by:
\begin{align}
    \frac{dp^r_1}{dt} &= -(\delta + \mu) p^r_1 - \ln \big( (1 - p_\text{rad})^{-1} \big) \sum^n_{j=1} \delta_D(t-s_j) p^r_1 \label{rad_deq_L1} \\
    \frac{dp^r_m}{dt} &= -(\delta + \mu) p^r_m + \delta p^r_{m-1} - \ln \big( (1 - p_\text{rad})^{-1} \big) \sum^n_{j=1} \delta_D(t-s_j) p^r_m, \, m \in [2, k] \label{rad_deq_Li} \\
    \frac{dp^r_{NL}}{dt} &= -(\alpha + \mu) p^r_{NL} + \delta p^r_k - \ln \big( (1 - p_\text{rad})^{-1} \big) \sum^n_{j=1} \delta_D(t-s_j) p^r_{NL} \label{rad_deq_NL} \\
    \frac{dp^r_{A}}{dt} &= - \gamma p^r_A + \alpha p^r_{NL} - \ln \big( (1 - p_\text{blood})^{-1} \big) \sum^n_{j=1} \delta_D(t-s_j) p^r_A \label{rad_deq_A} \\
    \frac{dp^r_C}{dt} &= \gamma p^r_A(t) + \ln \big( (1 - p_\text{blood})^{-1} \big) \sum^n_{j=1} \delta_D(t-s_j) p^r_A\\
    \frac{dp^r_D}{dt} &= \mu \bigg( \sum^{k}_{i=1} p^r_i + p^r_{NL} \bigg) + \ln \big( (1 - p_\text{rad})^{-1} \big) \sum^n_{j=1} \delta_D(t-s_j) \bigg( \sum^{k}_{i=1} p^r_i + p^r_{NL} \bigg)\label{rad_deq_D},
\end{align}
where $\delta_D(\cdot)$ denotes the Dirac delta function (not to be confused with $\delta$, a scalar parameter that denotes the rate of transition between successive latency compartments).\\

Here, we restrict our attention to a single administration of drug treatment at time $s_1$. For $t<s_1$, we note that $\mathbf{p^r}(t, s_1)=\mathbf{p}(t)$, as per Equations (\ref{l_eq}) to (\ref{d_eq}). Since we model the effects of radical cure to be instantaneous, $\mathbf{p^r}(t, s_1)$ will exhibit jump discontinuities at time $t=s_1$.\\

Integrating by parts, we can solve Equations (\ref{rad_deq_L1}) to (\ref{rad_deq_D}) for $t \geq s_1$ to obtain the state probabilities $\mathbf{p^r}(t, s_1)$ in terms of the state probabilities $\mathbf{p}(t)$ given by Equations (\ref{l_eq}) to (\ref{d_eq}):
\begin{align}
	p^r_{m}(t, s_1) =& \underbrace{(1-p_\text{rad}) \cdot p_m(t)}_{\substack{\text{radical cure failure: hypnozoite} \\ \text{(dormant) survives treatment at $s_1$}}} \text{ for } m \in [1, k] \label{l_rad_eq}\\
	p^r_{NL}(t, s_1) =& \underbrace{(1-p_\text{rad}) \cdot p_{NL}(t)}_{\substack{\text{radical cure failure: hypnozoite} \\ \text{(non-latent) survives treatment at $s_1$}}} \label{nl_rad_eq} \\
	p^r_A(t, s_1) =& \underbrace{(1-p_\text{blood}) e^{-\gamma(t-s_1)} p_A(s_1)}_{\substack{\text{blood-stage treatment failure:} \\ \text{drug fails to clear infection at $s_1$}}}  + \underbrace{(1-p_\text{rad}) \big( p_A(t) - e^{-\gamma(t-s_1)} p_A(s_1) \big)}_{\substack{\text{relapse triggered by a hypnozoite that survives} \\ \text{drug treatment at $s_1$ (i.e. radical cure failure)}}} \label{a_rad_eq} \\
	p^r_C(t, s_1) =& \underbrace{p_C(s_1)}_{\substack{\text{relapse cleared} \\ \text{naturally} \\ \text{before $s_1$}}} + \underbrace{p_\text{blood} \cdot p_A(s_1) }_{\substack{\text{blood-stage treatment success:} \\ \text{drug instantaneously clears} \\ \text{ongoing infection at $s_1$}}} + \underbrace{(1-p_\text{blood}) \big( 1 - e^{-\gamma(t-s_1) }\big) p_A(s_1)}_{\substack{\text{relapse surviving drug treatment at $s_1$} \\ \text{(i.e. blood-stage treatment failure)} \\ \text{is later cleared naturally} }} + \notag \\
	& \underbrace{(1-p_\text{rad}) \big[ p_C(t) - p_C(s_1) - ( 1 - e^{-\gamma(t-s_1)} p_A(s_1) ) \big]}_{\substack{\text{relapse triggered by a hypnozoite that survives} \\ \text{drug treatment at $s_1$ (i.e. radical cure failure)} \\ \text{is later cleared naturally}}} \label{c_rad_eq}\\
   p_D^r(t, s_1) =& \underbrace{p_D(s_1)}_{\substack{\text{hypnozoite dies} \\ \text{naturally} \\ \text{before $s_1$}}} + \underbrace{p_\text{rad} \cdot \Big( \sum^k_{m=1} p_m(s_1) +  p_{NL}(s_1) \Big) }_{\substack{\text{radical cure success: hypnozoite} \\ \text{(dormant or non-latent) instantaneously} \\ \text{killed by drug at $s_1$}}} + \underbrace{(1 - p_\text{rad}) [ p_D(t) - p_D(s_1) ]}_{\substack{\text{hypnozoite surviving treatment at $s_1$} \\ \text{(i.e. radical cure failure)} \\ \text{later dies naturally}}}. \label{d_rad_eq}
\end{align}

\begin{figure}
    \centering
    \includegraphics[width=\textwidth]{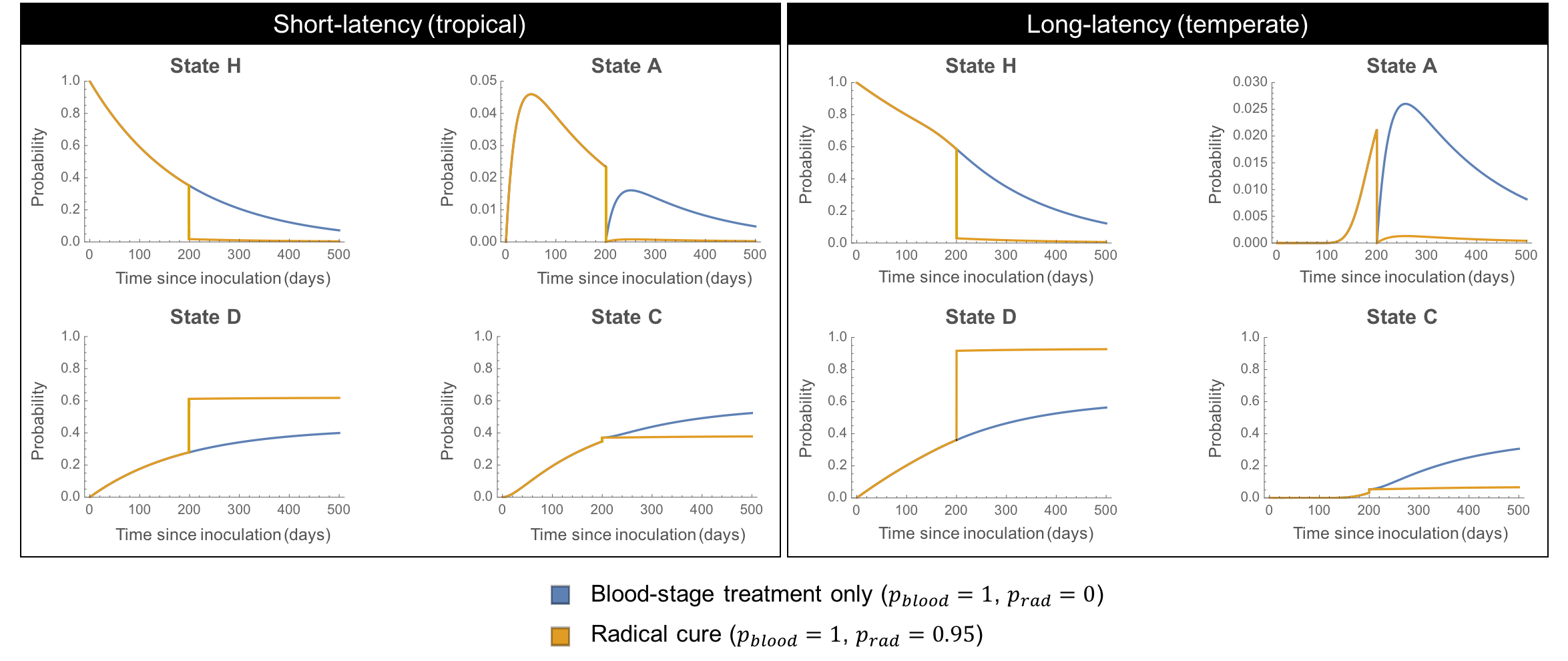}
    \caption{Activation-clearance dynamics, accounting for drug treatment at time $s_1=200$ days after inoculation, for both short-latency and long-latency hypnozoites using biologically-plausible parameter values. Baseline activation and clearance rates $\alpha=1/334$ day$^{-1}$ and $\mu=1/442$ day$^{-1}$, as well as the rate of progression through successive latency compartments $\delta=1/5$ day$^{-1}$ and the number of latency compartments $k=35$ have been obtained from \textcite{white2014modelling}. We further assume a clearance rate $\gamma=1/20$ day$^{-1}$ for blood-stage infection. We show the case of $p_\text{blood}=1$, $p_\text{rad}=0$ (blood-stage treatment only) and $p_\text{blood}=1$, $p_\text{rad}=0.95$ (reasonably effective radical cure).}
    \label{fig:single_hyp_illustrative}
\end{figure}

Illustrative results for our activation-clearance model, comparing blood-stage treatment only ($p_\text{blood}=1$, $p_\text{rad}=0$) against reasonably efficacious radical cure ($p_\text{blood}=1$, $p_\text{rad}=0.95$) are shown in Figure \ref{fig:single_hyp_illustrative} for both short-latency (tropical) and long-latency (temperate) strains. For long-latency strains, due to the enforced dormancy period, there is a delay of approximately $100$ days, during which a hypnozoite may die, but is highly unlikely to activate; short-latency hypnozoites, in contrast, may activate immediately after they are established in the liver. Upon drug treatment $s_1=200$ days after inoculation, all active infections (state $A$) are modelled to clear (state $C$) instantaneously, while a hypnozoite in the liver (state $H$) is modelled to die instantaneously (state $D$) with probability $p_\text{rad}$, leading to jump discontinuities in the state probabilities at time $t=s_1$. Each hypnozoite has two possible end states: death prior to activation (state $D$), or clearance of the blood-stage infection triggered by activation (state $C$). The steady-state probability of hypnozoite activation, given by the limit $p_C(t)$ as $t \to \infty$, is higher in the absence of radical cure. 

\section{Hypnozoite and Infection Dynamics in a General Transmission Setting} \label{sec::inf_server_queue}
We will now embed our relapse-clearance model in an epidemiological framework accounting for repeated mosquito inoculation. In previous work, we constructed an infinite server queue, with each departure corresponding to a hypnozoite activation event, to examine the cumulative number of relapses experienced in the interval $(0, t]$ in the absence of treatment, assuming an empty hypnozoite reservoir at time zero \parencite{mehra2021antibody}. Here, we construct an open \emph{network} of infinite server queues \parencite{harrison1981note} to jointly characterise at time $t$ the size of the hypnozoite reservoir; the number of ongoing relapses and primary infections; the number of relapses that have already been cleared; and the number of hypnozoites that have died prior to activation, whilst accounting for drug treatment (Section \ref{sec:queue_network}). We also extend our previous work to examine the cumulative number of blood-stage infections (that is, both primary infections and relapses) following drug treatment (Section \ref{sec::recur_after_drug}).

\subsection{Epidemiological Framework} \label{subsec:epi_framework}
We begin by extending the epidemiological framework introduced in \textcite{mehra2021antibody} to account for drug treatment and the dynamics of blood-stage infection. We assume that:
\begin{itemize}
    \item Infective mosquito bites follow a non-homogenous Poisson process with time-dependent rate $\lambda(t)$ such that the mean number of bites in the interval $(0, t]$, given by $m(t) = \int^t_0 \lambda(\tau) d\tau < \infty$ for all $t \geq 0$;
    \item Each mosquito bite establishes hypnozoites and, with probability $p_\text{prim}$, triggers a primary infection (state $P$), with independent dynamics for each bite;
    \item In the absence of treatment, blood-stage infections (primary and relapses) are cleared at rate $\gamma$ (that is, exponentially-distributed with expected duration $1/\gamma$);
    \item Any ongoing blood-stage infections (primary and relapses) are cleared instantaneously with probability $p_\text{blood}$ upon drug administration;
    \item The number of hypnozoites established by each mosquito bite is geometrically-distributed with mean $\nu$, as per \textcite{white2014modelling}; and
    \item Hypnozoite dynamics are independent and identically distributed (i.i.d.), with probability masses across states $1, \dots, k, NL, A, C, D$ given by
    \begin{itemize}
        \item Equations (\ref{l_eq}) to (\ref{d_eq}) in the absence of drug treatment
        \item Equations (\ref{l_rad_eq}) to (\ref{d_rad_eq}) given drug treatment is administered time $s_1$ after inoculation in the liver, where we assume that the drug instantaneously clears relapses (state $A$ to state $C$) with probability $p_\text{blood}$ and kills hypnozoites in the liver (states $1, \dots, k, NL$, collectively referred to as state $H$, to state $D$) with probability $p_\text{rad}$.
    \end{itemize}
\end{itemize}

\subsection{Open Network of Infinite Server Queues} \label{sec:queue_network}
\begin{figure}[h]
    \centering
    \includegraphics[width=\textwidth]{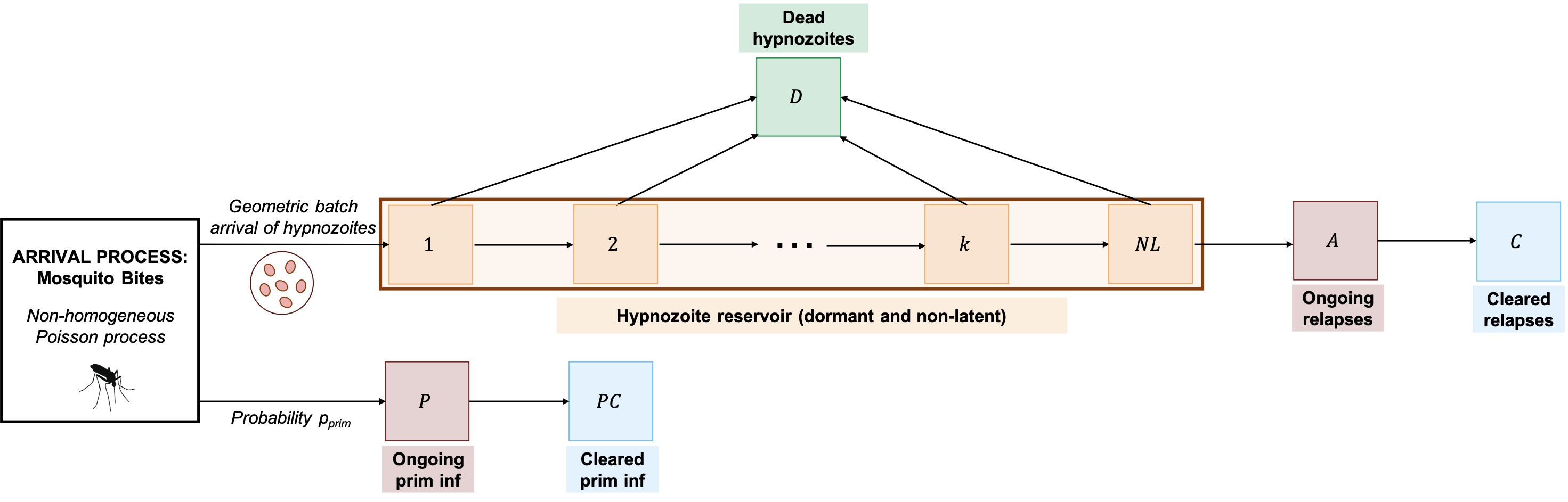}
    \caption{Schematic for open network of infinite server queues.}
    \label{fig:queue_network}
\end{figure}

To characterise hypnozoite and infection dynamics in a general transmission setting, we now construct an open network of infinite server queues, denoted $1, \dots, k, NL, A, D, C, P, PC$ (Figure \ref{fig:queue_network}). The arrival process for our network is comprised of mosquito bites, which we model as a non-homogeneous Poisson process with rate parameter $\lambda(t)$. Each mosquito bite is associated with a batch arrival (geometrically-distributed, with mean $\nu$) into either queue $1$ in the case of long-latency strains ($k>0$), or queue $NL$ in the case of short-latency strains ($k=0$), that is, a variable hypnozoite inoculum; and, with probability $p_\text{prim}$, a single arrival into queue $P$, that is, a primary infection.\\

For $m \in [1, k]$, in the absence of treatment, service times in queue $m$ are exponentially-distributed with rate $(\delta + \mu)$; a departure from queue $m$ may either be routed to queue $(m+1)$ (that is, the subsequent latency compartment) with probability $\delta/(\delta + \mu)$, or queue $D$ (that is, die due to the death of the host cell) with probability $\mu/(\delta + \mu)$. In contrast, service times in queue $NL$ are exponentially-distributed with parameter $\alpha + \mu$ in the absence of treatment; departures from queue $NL$ (that is, hypnozoites that have been cleared from the liver) are either routed to queue $D$ with probability $\mu/(\alpha + \mu)$, where they remain indefinitely (corresponding to hypnozoites that die prior to activation); or queue $A$ with probability $\alpha/(\alpha+\mu)$ (having activated to trigger a relapse). Upon the administration of drug treatment, with probability $p_\text{rad}$, each hypnozoite in queue $1, \dots, k, NL$ is immediately routed to queue $D$ (that is, killed due to radical cure).\\

Service times in both queues $A$ and $P$, which correspond to the duration of relapses and primary infections respectively, are exponentially-distributed with rate $\gamma$ in the absence of treatment. Upon the administration of drug treatment, with probability $p_\text{blood}$, each relapse immediately departs queue $A$, while each primary infection immediately departs queue $P$. Departures from queues $A$ and $P$ (that is, cleared relapses and primary infections) are routed to queues $C$ and $PC$ respectively, where they remain indefinitely.\\

For a single hypnozoite that enters either queue $1$ in the case of long-latency strains ($k>0$) or queue $NL$ in the case of short-latency strains ($k=0$), our network of queues captures precisely those dynamics discussed in Section \ref{sec::single_hyp}, with the probability of the hypnozoite being present in queue $s \in \{1, \dots, k, NL, A, C, D \}$ at time $t$ after inoculation (that is, arrival into the network) given by the state probabilities (\ref{l_rad_eq}) to (\ref{d_rad_eq}). Upon inoculation, we assume that each hypnozoite and infection behaves independently \parencite{harrison1981note}.\\ 

Suppose an individual is first exposed to infective mosquito bites at time zero. Let $N_s(t)$ denote the number of hypnozoites/infections at time $t$ in queue $s \in \{ 1, \dots, k, NL, A, C, D, P \} =: S$. We seek to derive a probability generating function (PGF) for the random vector
\begin{align*}
    \mathbf{N}(t) = (N_1(t), N_2(t), \dots, N_k(t), N_{NL}(t), N_A(t), N_C(t), N_D(t), N_P(t), N_{PC}(t)).
\end{align*}

At time zero, we assume an empty hypnozoite reservoir with no prior infection history, that is,
\begin{align*}
    \mathbf{N}(0) = \mathbf{0}.
\end{align*}

For notational convenience, we introduce a superscript $\mathbf{N}^{t_1}(t)$ to capture the administration of radical cure at time $t_1$; the absence of a superscript indicates a scenario with no drug treatment.\\

We analyse this network of queues by first considering the case of a single arrival event (mosquito bite). In Section \ref{sec::baseline_bite}, we examine hypnozoite and infection dynamics for a single mosquito bite in the absence of treatment; we extend this analysis to account for drug treatment in Section \ref{sec::rc_queue}. By examining the properties of the non-homogeneous Poisson process governing mosquito bites, we obtain a PGF for $\mathbf{N}(t)$ in Section \ref{sec::baseline_inoc}.

\subsubsection{Dynamics for a Single Bite in the Absence of Treatment} \label{sec::baseline_bite}
Here we consider a single mosquito bite at time $\tau$ in the absence of drug treatment. To characterise infection and hypnozoite dynamics arising from the bite, we condition on the size of the hypnozoite inoculum.\\

We begin by examining hypnozoites and relapses only. Suppose a single hypnozoite is established in the liver (that is, enters the network of queues) at time $\tau$. Then the joint PGF for the number of hypnozoites in each queue $S_h = \{ 1, \dots, k, NL, A, C, D \}$ at the $t \geq \tau$ follows readily from the state probabilities for a single hypnozoite, for which we have analytic solutions given by Equations (\ref{l_eq}) to (\ref{d_eq})
\begin{align*}
    \EX \Big[ \prod_{s \in S_h} z_s^{N_s(t)} | \, 1 \text{  hypnozoite at time } \tau \Big] = \sum_{s \in S_h} z_s \cdot p_s(t-\tau).    
\end{align*}

Now, suppose a bite establishing exactly $n$ hypnozoites occurs at time $\tau$, that is, a batch of $n$ hypnozoites enters the queue at time $\tau$. Assuming that hypnozoite dynamics are i.i.d., it follows that
\begin{align*}
    \EX \Big[ \prod_{s \in S_h} z_s^{N_s(t)} | \, n \text{ hypnozoites at time } \tau \Big] = \Big( \sum_{s \in S_h} z_s \cdot p_s(t-\tau) \Big)^n.
\end{align*}

Given the number of hypnozoites established by each mosquito bite is geometrically-distributed with mean $\nu$, by the law of total expectation,
\begin{align}
    \EX \Big[ \prod_{s \in S_h} z_s^{N_s(t)} | \text{ bite at time } \tau \Big] & = \sum^\infty_{n=0}  \frac{1}{\nu + 1} \Big( \frac{\nu}{\nu + 1} \Big)^n \cdot \EX \Big[ \prod_{s \in S_h} z_s^{N_s(t)} | \, n \text{ hypnozoites at time } \tau \Big] \notag \\ 
    & = \frac{1}{\nu + 1} \sum^\infty_{n=0} \Big( \sum_{s \in S_h} z_s \cdot p_s(t-\tau) \Big)^n \Big( \frac{\nu}{1+\nu} \Big)^n \notag \\
    & = \frac{1}{1 + \nu \Big( 1- \sum_{s \in S_h} z_s \cdot p_s(t-\tau) \Big)}, \label{hyp_bite_pgf}
\end{align}
where the geometric series summation converges in the domain $|z_s| \leq 1$ for all $s \in S_h$.\\

Recall that each mosquito bite is assumed to trigger a primary infection with probability $p_\text{prim}$, with primary infections cleared at rate $\gamma$. The joint PGF for the number of ongoing ($N_P(t)$) and cleared ($N_{PC}(t)$) primary infections at time $t \geq \tau$ arising from the bite is therefore
\begin{align}
    \EX \big[ z_{P}^{N_P(t)} z_{PC}^{N_{PC}(t)} | \text{ bite at time } \tau \big] =  \underbrace{(1-p_\text{prim})}_{\substack{\text{no prim inf} \\ \text{ due to bite}}} + \underbrace{p_\text{prim} e^{-\gamma(t-\tau)}}_{\substack{\text{ongoing prim inf} \\ \text{at time $t$}}} z_P + \underbrace{p_\text{prim} (1 -  e^{-\gamma (t-\tau)}) z_{PC}}_{\substack{\text{prim inf cleared by time $t$}}}. \label{prim_bite_pgf}
\end{align}

Given primary infection dynamics are independent of hypnozoite and relapse dynamics, it follows from Equation (\ref{hyp_bite_pgf}) (the joint PGF for $(N_1(t), N_2(t), \dots, N_k(t), N_{NL}(t), N_{A}(t), N_C(t), N_D(t))$) and Equation (\ref{prim_bite_pgf}) (the joint PGF for $(N_P(t), N_{PC}(t))$) that
\begin{align}
    \EX \Big[ \prod_{s \in S} z_s^{N_s(t)} | \text{ bite at time } \tau \Big]  &= \EX \Big[ \prod_{s \in S_h} z_s^{N_s(t)} | \text{ bite at time } \tau \Big] \cdot \EX \big[ z_{P}^{N_P(t)} z_{PC}^{N_{PC}(t)} | \text{ bite at time } \tau \big] \notag \\
    &= \frac{p_\text{prim} e^{-\gamma (t-\tau)} z_P + p_\text{prim}(1-  e^{-\gamma (t-\tau)}) z_{PC} + (1-p_\text{prim})}{1 + \nu \Big( 1- \sum_{s \in S_h} z_s \cdot p_s(t-\tau) \Big)}. \label{baseline_bite_pgf}
\end{align}
We note that Equation (\ref{baseline_bite_pgf}) holds in the domain $|z_s| \leq 1$ for each $s \in S$.\\

Given a single mosquito bite at time $\tau \leq t$, Equation (\ref{baseline_bite_pgf}) characterises the joint PGF for the number of hypnozoites/infections in each queue $S \in \{ 1, \dots, k, NL, A, C, D, P, PC \}$ at time $t$ in the absence of treatment. 

\subsubsection{Dynamics for a Single Bite Under Drug Treatment} \label{sec::rc_queue}
Now, suppose that drug treatment is administered at time $t_1$, instantaneously killing each hypnozoite in the liver with probability $p_\text{rad}$ and clearing each ongoing blood-stage infection (primary or relapse) with probability $p_\text{blood}$. Here, we consider the case $t \geq t_1$.\\

As in Section \ref{sec::baseline_bite}, we begin by considering a single hypnozoite that is established in the liver at time $\tau$ to obtain
\begin{align*}
    \EX \Big[ \prod_{s \in S_h} z_s^{N^{t_1}_s(t)} | \, 1 \text{  hypnozoite at time } \tau \Big] = \sum_{s \in S_h} z_s \cdot p^r_s(t-\tau, t_1 - \tau),
\end{align*}
where the probability $p^r_s(t-\tau, t_1 - \tau)$ of a hypnozoite being in queue $s \in S_h = \{1, \dots, k, NL, A, C, D \}$ at time $t \geq t_1$, accounting for drug treatment at time $t_1$, is given by Equations (\ref{l_rad_eq}) to (\ref{d_rad_eq}).\\

Assuming that hypnozoite dynamics are i.i.d. and hypnozoite inocula are geometrically distributed with mean $\nu$, using similar reasoning to Section \ref{sec::baseline_bite}, it follows that
\begin{align}
    \EX \Big[ \prod_{s \in S_h} z_s^{N^{t_1}_s(t)} | \text{ bite at time } \tau \Big] = \frac{1}{1 + \nu \Big( 1- \sum_{s \in S_h} z_s \cdot p^r_s(t-\tau, t_1 - \tau) \Big)}, \label{drug_hyp_bite_pgf}
\end{align}
where the RHS is well-defined in the domain $|z_s| \leq 1$ for each $s \in S_h$. Equation (\ref{drug_hyp_bite_pgf}) characterises hypnozoite and relapse dynamics for a single bite following drug treatment. We next account for primary infections, which occur with probability $p_\text{prim}$ for each mosquito bite. In the case of a bite that occurs after drug treatment, that is, $\tau \geq t_1$, primary infections are unaffected by drug treatment, hence
\begin{align*}
    \EX \Big[ z_P^{N^{t_1}_p(t)} z_{PC}^{N^{t_1}_{pc}(t)} | \text{ bite at time } \tau \geq t_1 \Big] = (1-p_\text{prim}) + p_\text{prim} e^{-\gamma(t-\tau)} z_P + p_\text{prim} (1 -  e^{-\gamma (t-\tau)}) z_{PC}
\end{align*}
as per Equation (\ref{prim_bite_pgf}). For bites prior to drug treatment, that is $\tau < t_1$, the joint PGF for the number of ongoing ($N^{t_1}_P(t)$) and cleared ($N^{t_1}_{PC}(t)$) primary infections at time $t \geq t_1$ is given by
\begin{align}
    \EX \Big[ & z_P^{N^{t_1}_p(t)} z_{PC}^{N^{t_1}_{pc}(t)}  | \text{ bite at time } \tau < t_1 \Big] =  \underbrace{(1-p_\text{prim})}_{\substack{\text{no prim inf} \\ \text{ due to bite}}} + \underbrace{p_\text{prim} (1 - p_\text{blood}) e^{-\gamma(t-\tau)}}_{\substack{\text{ongoing prim inf at time $t \geq t_1$} \\ \text{that survives drug at time $t_1$}}} z_P + \notag \\ 
    & \bigg( \underbrace{p_\text{prim} p_\text{blood} e^{-\gamma (t_1-\tau)} }_{\substack{\text{prim inf cleared} \\ \text{due to drug at time $t_1$}}}  + \underbrace{p_\text{prim} (1-p_\text{blood}) e^{-\gamma (t_1-\tau)} [ 1 - e^{-\gamma(t-t_1)}] }_{\substack{\text{prim inf that survives drug at time $t_1$} \\ \text{cleared naturally by time $t$}}} + \underbrace{p_\text{prim} (1 - e^{-\gamma(t_1 - \tau)})}_{\substack{\text{prim inf cleared naturally} \\ \text{ before treatment at time $t_1$}}} \bigg) z_{PC}. \label{drug_prim_bite_pgf}
\end{align}

Assuming hypnozoite dynamics are independent to primary infections, for $t\geq t_1$, it follows from Equation (\ref{drug_hyp_bite_pgf}) (the joint PGF for $(N^{t_1}_1(t), N^{t_1}_2(t), \dots, N^{t_1}_k(t), N^{t_1}_{NL}(t), N^{t_1}_{A}(t), N^{t_1}_C(t), N^{t_1}_D(t))$) and Equations (\ref{prim_bite_pgf}) and (\ref{drug_prim_bite_pgf}) (the joint PGF for $(N^{t_1}_P(t), N^{t_1}_{PC}(t))$ in the cases $\tau \geq t_1$ and $\tau < t_1$ respectively) that
\begin{align}
    \EX \Big[ \prod_{s \in S} z_s^{N^{t_1}_s(t)} | &\text{ bite at time } \tau \Big]   = \EX \Big[ \prod_{s \in S_h} z_s^{N^{t_1}_s(t)} | \text{ bite at time } \tau \Big] \cdot  \EX \Big[ z_P^{N^{t_1}_p(t)} z_{PC}^{N^{t_1}_{pc}(t)}  | \text{ bite at time } \tau \Big] \notag \\
    & = \begin{cases}
      \frac{1-p_\text{prim} + p_\text{prim} e^{-\gamma(t-\tau)} z_P + p_\text{prim} (1 - e^{-\gamma(t-\tau)}) z_{PC} }{1 + \nu \big( 1- \sum_{s \in S_h} z_s \cdot p_s(t-\tau) \big)} & \text{ if } \tau \geq t_1\\
        \frac{1-p_\text{prim} + p_\text{prim} (1 - p_\text{blood}) e^{-\gamma(t-\tau)} z_P + p_\text{prim} (1 - (1 - p_\text{blood}) e^{-\gamma(t-\tau)}) z_{PC} }{1 + \nu \big( 1- \sum_{s \in S_h} z_s \cdot p^r_s(t-\tau, t_1 - \tau) \big)} & \text{ if } \tau < t_1
    \end{cases}
  . \label{rc_bite_pgf}
\end{align}
where $S_h = \{ 1, 2, \dots, k, NL, A, C, D \}$ denotes the state space for a single hypnozoite. We note that Equation (\ref{rc_bite_pgf}) holds in the domain $|z_s| \leq 1$ for each $s \in S$.\\

\begin{figure}[p]
    \centering
    \includegraphics[width=\textwidth]{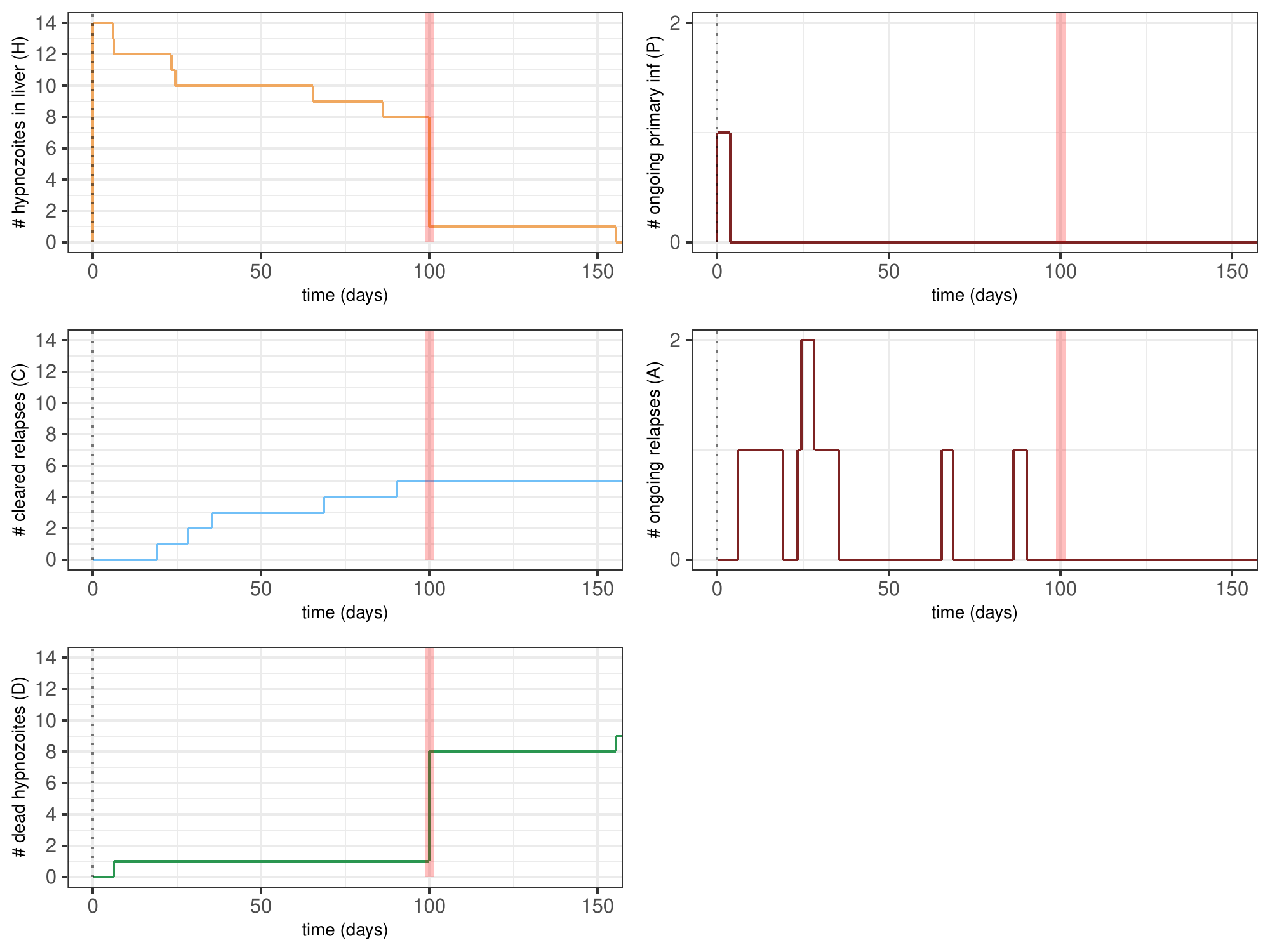}
    \caption{Simulated sample path for a single mosquito bite, assuming short-latency (tropical) strains. At time $t=0$, an infective mosquito bite triggers a primary infection and establishes $14$ hypnozoites in the host liver. We account for the administration of radical cure ($p_\text{rad}=0.95$, $p_\text{blood}=1$) at time $t=100$ days following the bite. Hypnozoite activation and death rates, $\alpha=1/334 \text{ day}^{-1}$ and $\mu=1/442 \text{ day}^{-1}$, are based on estimates by \textcite{white2014modelling}. The clearance rate for blood-stage infections has been set to $\gamma=1/20 \text{ day}^{-1}$.}
    \label{fig:sample_path_bite}
\end{figure}

For a single a mosquito bite at time $\tau \leq t$, the joint PGF in Equation (\ref{rc_bite_pgf}) characterises the number of hypnozoites/infections in each queue $S \in \{1, \dots, k, NL, A, C, D, P, PC \}$ at time $t \geq t_1$, given the administration of a drug at time $t_1$. An illustrative sample path, generated using direct stochastic simulation (using the Doob-Gillespie algorithm) of short-latency hypnozoite dynamics, is shown in Figure \ref{fig:sample_path_bite}. At time $t=0$, an infective mosquito bite establishes $14$ hypnozoites in the host liver, and triggers a primary infection; radical cure is administered $t=100$ days following the mosquito bite, as indicated by the vertical red line. Prior to $t=100$ days, the number of hypnozoites in the liver (state $H$) decreases over time as hypnozoites either activate to cause relapses (state $A$) or die prior to activation (state $D$) at constant rates. Hypnozoite activation events in quick succession lead to overlapping relapses for a brief period of time, with all relapses eventually cleared (state $C$). Upon the administration of radical cure (vertical red line), seven of the eight remaining hypnozoites within the liver are instantaneously killed. The final hypnozoite then dies, prior to activation, approximately $160$ days after the mosquito bite.

\subsubsection{Mosquito Inoculation} \label{sec::baseline_inoc}
We now consider the non-homogeneous Poisson process (of rate $\lambda(t)$) governing mosquito bites. To characterise the PGF for $\mathbf{N}(t)$ as a function of hypnozoite and infection dynamics for a single mosquito bite (as examined in Sections \ref{sec::baseline_bite} and \ref{sec::rc_queue}), we first condition on the number of mosquito bites in a given interval, and then the bite times themselves, following the procedure detailed in \textcite{mehra2021antibody} and based on \textcite{parzen1999stochastic}.\\

Let $M(t)$ denote the number of infective mosquito bites in the interval $(0, t]$, with respective bite times $T_i$, $i \in \{1, \dots, M(t) \}$. Applying the law of total expectation and recalling that $M(t) \sim \text{Poisson}(m(t))$, where $m(t) = \int^t_0 \lambda(\tau) d \tau$
denotes the mean number of bites in the interval $(0, t]$, we have
\begin{align}
    \EX \Big[ \prod_{s \in S} z_s^{N_s(t)} \Big] &= \sum^\infty_{m=0} \frac{m(t)^m e^{-m(t)}}{m!} \EX \Big[ \prod_{s \in S} z_s^{N_s(t)} | M(t) = m \Big]. \label{baseline_lte_nbite}
\end{align}
where $S=\{ 1, \dots, k, NL, A, C, D, P, PC \}$ denotes the set of queues in the network.\\

Now, suppose $M(t)=m$, that is, precisely $m$ bites occur in the interval $(0, t]$. Assuming that hypnozoite and infection dynamics arising from each mosquito bite are independent, we have
\begin{align}
    \EX \Big[ \prod_{s \in S} z_s^{N_s(t)} | M(t) = m, T_1 = \tau_1, \dots, T_m = \tau_m \Big] &= \prod^m_{j=1} \EX \Big[ \prod_{s \in S} z_s^{N_s(t)} | \text{ bite at time } \tau_j \Big]. \label{cond_exp_1}
\end{align}

As per \textcite{lewis1967non}, the bite times $T_1, \dots, T_m$ have a conditional distribution equivalent to $m$ i.i.d. random variables with density
\begin{align*}
    f(\tau) = \frac{\lambda(\tau)}{m(t)} \mathbbm{1}\{ \tau \in [0, t) \},
\end{align*}
and thus have joint density
\begin{align}
    f_{T_1, \dots, T_m}(\tau_1, \dots, \tau_m) = \prod^m_{j=1} f(\tau_j) = \frac{1}{m(t)^m} \prod^m_{j=1} \lambda(\tau_j) \mathbbm{1} \big( \tau_j \in (0, t] \big). \label{joint_dens_1}
\end{align}

By integrating the conditional expectation given by Equation (\ref{cond_exp_1}) over the joint density given by Equation (\ref{joint_dens_1}), we obtain 
\begin{align}
    \EX \Big[ &\prod_{s \in S} z_s^{N_s(t)} | M(t) = m \Big] \notag \\
    &= \int^\infty_0 d\tau_1 \dots \int^\infty_0 d\tau_m  \EX \Big[ \prod_{s \in S} z_s^{N_s(t)} | M(t) = m, T_1 = \tau_1, \dots, T_m = \tau_m \Big] f_{T_1, \dots, T_m}(\tau_1, \dots, \tau_m) \notag \\
    &= \frac{1}{m(t)^m} \int^t_0 d\tau_1 \dots \int^t_0 d\tau_m \prod^m_{j=1} \Big\{ \lambda(\tau_j) \EX \Big[ \prod_{s \in S} z_s^{N_s(t)} | \text{ bite at time } \tau_j \Big] \Big\} \notag \\
    & = \Bigg( \frac{1}{m(t)} \int^t_0 \lambda(\tau) \EX \Big[ \prod_{s \in S} z_s^{N_s(t)} | \text{ bite at time } \tau \Big] d \tau \Bigg)^m. \label{baseline_pgf_nbite}
\end{align}

Substituting Equation (\ref{baseline_pgf_nbite}) into Equation (\ref{baseline_lte_nbite}) yields the joint PGF for $\mathbf{N}(t)$ in a general transmission setting as a function of the PGF for a single mosquito bite:
\begin{align}
     \EX \Big[ \prod_{s \in S} z_s^{N_s(t)} \Big] &= e^{-m(t)} \sum_{m=0}^\infty  \frac{m(t)^m}{m!} \Bigg[ \frac{1}{m(t)} \int^t_0 \lambda(\tau) \EX \Big[ \prod_{s \in S} z_s^{N_s(t)} | \text{ bite at time } \tau \Big] d \tau \Bigg]^m \notag \\
    & = \exp \bigg\{ -m(t) + \int^t_0 \lambda(\tau) \EX \Big[ \prod_{s \in S} z_s^{N_s(t)} | \text{ bite at time } \tau \Big] d \tau \bigg\} \label{multi_pgf_general},
\end{align}
where the RHS has been simplified using the Taylor series expansion of the exponential function.\\

To capture hypnozoite and infection dynamics in the absence of drug treatment, we substitute Equation (\ref{baseline_bite_pgf}) into Equation (\ref{multi_pgf_general}) to yield the joint PGF for $\mathbf{N}(t)$,
\begin{align}
     G(t, z_1, & \dots,  z_k, z_{NL}, z_A, z_D, z_C, z_P, z_{PC}) := \EX \Big[ \prod_{s \in S} z_s^{N_s(t)} \Big] \notag\\ 
     & = \exp \bigg\{ -m(t) + \int^t_0 \lambda(\tau) \frac{1-p_\text{prim} + p_\text{prim} e^{-\gamma (t-\tau)} z_P + p_\text{prim}(1-  e^{-\gamma (t-\tau)}) z_{PC}}{1 + \nu \Big( 1- \sum_{s \in S_h} z_s \cdot p_s(t-\tau) \Big)} d \tau \bigg\}. \label{multi_pgf}
\end{align}

In the case where drug treatment is administered at time $t_1$, we obtain the joint PGF for $\mathbf{N}^{t_1}(t)$ by substituting Equation (\ref{rc_bite_pgf}) in Equation (\ref{baseline_pgf_nbite}):
\begin{align}
     G^{t_1}(t, z_1, & \dots,  z_k, z_{NL}, z_A, z_D, z_C, z_P, z_{PC}) := \EX \Big[ \prod_{s \in S} z_s^{N^{t_1}_s(t)} \Big] \notag \\
      = \exp \bigg\{ & -m(t) +  \int^t_{t_1} \lambda(\tau) \frac{1-p_\text{prim} + p_\text{prim} e^{-\gamma (t-\tau)} z_P + p_\text{prim}(1-  e^{-\gamma (t-\tau)}) z_{PC}}{1 + \nu \Big( 1- \sum_{s \in S_h} z_s \cdot p_s(t-\tau) \Big)} d \tau + \notag\\
     & \int^{t_1}_0 \lambda(\tau ) \frac{ 1-p_\text{prim} + p_\text{prim} (1-p_\text{blood}) e^{-\gamma (t-\tau)} z_P + p_\text{prim}(1-  (1-p_\text{blood}) e^{-\gamma (t-\tau)}) z_{PC} }{1 + \nu \Big( 1- \sum_{s \in S_h} z_s \cdot p^r_s(t-\tau, t_1 - \tau) \Big)} d \tau \bigg\} \label{multi_rad_pgf}.
\end{align}

Both Equations (\ref{multi_pgf}) and (\ref{multi_rad_pgf}) are well-defined in the domain $|z_s| \leq 1$ for each $s \in S$. Here, we recall that
\begin{itemize}
    \item Mosquito bites follow a non-homogeneous Poisson process with rate $\lambda(t)$, with the mean number of bites in the interval $(0, t]$ given by $m(t) = \int^{t}_0 \lambda(\tau) d \tau$;
    \item Each bite triggers a primary infection with probability $p_\text{prim}$ and establishes geometrically-distributed hypnozoite inocula with mean $\nu$;
    \item Blood-stage infections (primary or relapse) are cleared at rate $\gamma$ at baseline, but are cleared instantaneously with probability $p_\text{blood}$ upon treatment at time $t_1$;
    \item State probabilities for a single hypnozoite, with state space $S_h = \{ 1, \dots, k, NL, A, C, D \}$ are given by Equations (\ref{l_eq}) to (\ref{d_eq}) in the absence of treatment ($p_s$) and Equations (\ref{l_rad_eq}) to (\ref{d_rad_eq}) following drug treatment at time $t_1$ ($p^r_s$), with the latter state probabilities accounting for each hypnozoite in the liver (states $1, \dots, k, NL$) being killed instantaneously (state $D$) with probability $p_\text{rad}$ upon treatment.
\end{itemize}

\begin{figure}
    \centering
    \includegraphics[width=\textwidth]{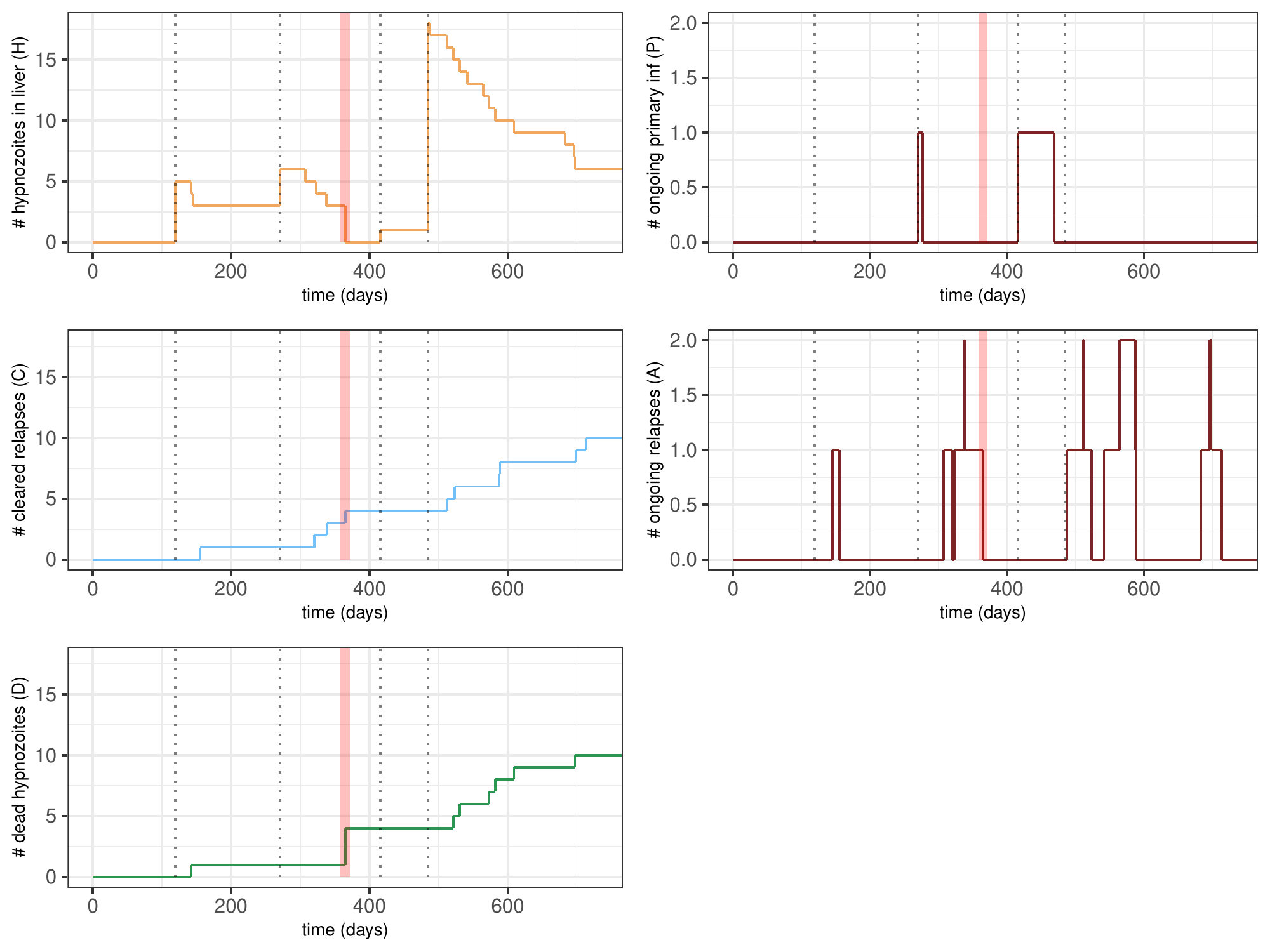}
    \caption{Simulated infection and hypnozoite dynamics for an individual in a constant transmission setting, assuming short-latency (tropical) strains. At time $t=0$, we assume an empty hypnozoite reservoir, with no ongoing infections. Radical cure ($p_\text{rad}=0.95$, $p_\text{blood}=1$) is adminstered after the hypnozoite reservoir has accumulated for $t=365$ days. In this simulation, an individual is bitten four times over a two year period, as indicated with dashed vertical lines. Mosquito bites have been modelled to follow a Poisson process with constant rate $\lambda=3/365 \text{ day}^{-1}$, with each bite establishing an average of $\nu=9$ hypnozoites in the liver (as per estimates from \textcite{white2014modelling}) and triggering a relapse with probability $p_\text{prim}=0.5$. Hypnozoite activation and death rates, $\alpha=1/334 \text{ day}^{-1}$ and $\mu=1/442 \text{ day}^{-1}$, are based on estimates by \textcite{white2014modelling}. The clearance rate for blood-stage infections has been set to $\gamma=1/20 \text{ day}^{-1}$.}
    \label{fig:sample_path_epi}
\end{figure}

An illustrative sample path of infection dynamics for an individual in a constant transmission setting is shown in Figure \ref{fig:sample_path_epi}, assuming short-latency (tropical) strains. Activation-clearance dynamics for each hypnozoite have been simulated using direct stochastic simulation (using the Doob-Gillespie algorithm). At time $t=0$, we assume that the individual has both an empty hypnozoite reservoir and no ongoing infections. In this simulation, an individual receives four infective bites (as indicated with dashed vertical lines) over a two-year period, with two bites triggering primary infections (state $P$). The hypnozoite reservoir (state $H$) fluctuates in size as hypnozoites are replenished through infective mosquito bites, but removed from the liver, either due to activation, thereby triggering relapses (state $A$), or death (state $D$). Hypnozoite activation events in quick succession give rise to overlapping relapses (that is, multiple infections). Radical cure, administered after the hypnozoite reservoir has been allowed to accumulate for $t=365$ days (indicated with a vertical red line), kills the entire hypnozoite reservoir and instantaneously clears an ongoing relapse. 

\subsection{Recurrences Following Drug Treatment} \label{sec::recur_after_drug}
In Sections \ref{sec::baseline_bite} and \ref{sec::rc_queue}, we examined the dynamics of the hypnozoite reservoir in an interval $(0, t]$, where time zero marks the time of first exposure in the epidemiological setting. Here, under the same epidemiological framework, we instead consider the cumulative number of recurrences in the interval $(t_1, t_2]$ following drug treatment at time $t_1$. Denoting
\begin{align*}
    I_C(t) = N_A(t) + N_C(t) + N_P(t) + N_{PC}(t),
\end{align*}
we seek to derive a PGF for the quantity $I_C(t_2) - I_C(t_1)$. The number of recurrences following drug treatment can provide insight into the impact of drug treatment on the infection burden. Since the random variables $I_C(t_2) - I_C(t_1)$ and $I_C(t_1)$ are not independent, we cannot characterise the quantity $I_C(t_2)-I_C(t_1)$ directly from the results in Section \ref{sec:queue_network}.\\

\begin{figure}[p]
    \centering
    \includegraphics[width=\textwidth]{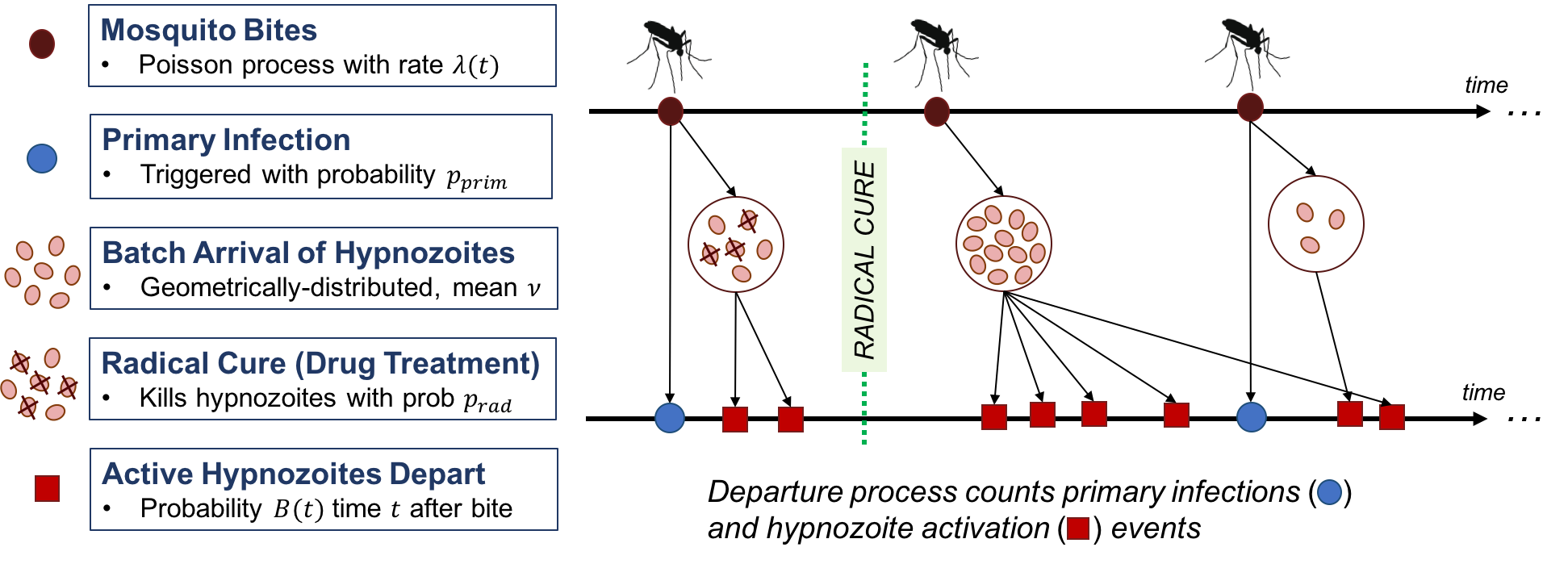}
    \caption{Schematic for model of \textit{P. vivax} recurrences. To capture recurrences due to both reinfection and hypnozoite activation, we construct an infinite server queue. Arrivals, which represent mosquito bites, occur according to a non-homogeneous Poisson process with time-dependent rate $\lambda(t)$. Each mosquito bite triggers a primary infection with probability $p_\text{prim}$, and also leads to the establishment of hypnozoites in the liver. The hypnozoite inoculum associated with each bite is assumed geometrically-distributed with mean $\nu$. The service time of each hypnozoite is i.i.d. with distribution $B(t)=p_A(t) + p_C(t)$ (Equation (\ref{b_eq}), which describes the probability that a hypnozoite has activated time $t$ after inoculation, noting that not every hypnozoite necessarily activates (that is, it is not guaranteed that $\lim_{t \to \infty} B(t) = 1$). Under this formulation, only active hypnozoites and primary infections depart the queue. The departure process counts the cumulative number of recurrences over time. To model the effects of radical cure, we assume that any hypnozoite established before drug treatment is killed instantaneously with probability $p_\text{rad}$ upon treatment.}
    \label{fig:schematic}
\end{figure}

Similarly to \textcite{mehra2021antibody}, we now construct an infinite server queue, such that the departure process counts the cumulative number of recurrences over time (Figure \ref{fig:schematic}). Noting that the departure process of an infinite server queue constitutes a shot noise process \parencite{holman1983service}, the total number of recurrences in the interval $[0, t_2)$, $t_2 \geq t_1$ can be written
\begin{align*}
    I_C(t_2) = \underbrace{\sum^{N(t_1)}_{m=1} \big( M^r_m(t_2-\tau_m) + V_m \big)}_{\text{\# recurrences from bites in $(0, t_1]$}} + \underbrace{\sum^{N(t_2)}_{m=N(t_1) + 1} \big( M_m(t_2-\tau_m) + V_m \big)}_{\text{\# recurrences from bites in $(t_1, t_2]$}},
\end{align*}
where
\begin{itemize}
    \item $N(t) \sim \text{Poisson}(m(t))$ denotes the number of mosquito bites in the interval $[0, t)$,  with $\tau_1, \dots, \tau_{N(t)}$ denoting the respective bite times;
    \item $M^r_m(t_2-\tau_m)$ are independent random variables, representing the number of hypnozoite activation events at time $t_2$ arising from a bite that occurred at time $\tau_m < t_1$ whilst accounting for drug treatment at time $t_1$
    \item $M_m(t_2-\tau_m)$ are independent random variables representing the number of hypnozoite activation events at time $t_2$ arising from a bite that occurred at time $\tau_m > t_1$ and are thus unaffected by drug treatment; and
    \item $V_m \overset{iid}{\sim} \text{Bernoulli}(p_\text{prim})$ represent primary infections, which occur independently with probability $p_\text{prim}$ as a result of each mosquito bite.
\end{itemize}

We seek to characterise the number of recurrences in the interval $(t_1, t_2]$, that is,
\begin{align}
    I_C(t_2) - I_C(t_1) &= \underbrace{\sum^{N(t_1 )}_{m=0} \big( M^r_m(t_2-\tau_m) - M^r_m(t_1-\tau_m) \big)}_{\substack{\text{\# recurrences initiated in $(t_1, t_2]$} \\ \text{from bites in $(0, t_1]$}}} + \underbrace{\sum^{N(t_2)}_{m=N(t_1)+1} \big( M_m(t_2-\tau_m) + V_m \big)}_{\substack{\text{\# recurrences initiated in $(t_1, t_2]$} \\ \text{from bites in $(t_1, t_2]$}}} \notag \\
    &=: I_{C_1}(t_1, t_2) + I_{C_2}(t_1, t_2). \label{ic_prelim}
\end{align}
Only hypnozoites and infections established before time $t_1$ are affected by drug treatment; for all bites following drug treatment, we revert to the scenario detailed in Section \ref{sec::baseline_bite}, which does not consider drug treatment. As such, setting $t_1=0$ in Equation (\ref{ic_prelim}) yields the cumulative number of infections in the interval $(0, t_2]$, where zero represents the time since an individual is first exposed to infective mosquito bites, in the absence of treatment.\\

By the independent increment property of the Poisson process, the number of bites in the disjoint intervals
\begin{align*}
    &(0, t_1]: \, N(t_1) \sim \text{Poisson}(m(t_1 ))  \\
    &(t_1, t_2]: \, N(t_2) - N(t_1 )  \sim \text{Poisson}(m(t_2) -m( t_1 )) 
\end{align*}
are independent random variables. The bite times in the intervals $(0, t_1]$ and $(t_1, t_2]$ are also independent. Given the dynamics of each hypnozoite and primary infection are independent, it follows that $I_{C_1}(t_1,t_2)$ and $I_{C_1}(t_1,t_2)$, the number of recurrences triggered in $(t_1, t_2]$ by bites in the intervals $(0, t_1]$ and $(t_1, t_2]$ respectively, are independent. Thus, from Equation (\ref{ic_prelim}), the PGF for $I_C(t_2) - I_C(t_1)$ is given by the product
\begin{align}
    \EX \big[ z^{I_C(t_2) - I_C(t_1)} \big] = \EX \big[ z^{I_{C_1}(t_1, t_2)} \big] \EX \big[ z^{I_{C_2}(t_1, t_2)} \big]. \label{ic_prod}
\end{align}
We therefore proceed by considering mosquito bites in the intervals $(0, t_1]$ and $(t_1, t_2]$ separately.\\

We begin by analysing recurrences arising from a single mosquito bite at time $\tau_m \in (t_1, t_2]$, that is, the random variable $(M_m(t_2 - \tau_m) + V_m)$. Hypnozoites and infections triggered by this bite will \emph{not} be affected by drug treatment.\\

Suppose a single hypnozoite is established in the liver at time $\tau_m \in (t_1, t_2]$. For notational convenience, we denote the probability this hypnozoite will activate in the interval $(\tau_m, t_2]$ by $B(t_2 - \tau_m)$, where
\begin{align}
    B(t) & =  p_A(t) + p_C(t) \notag  \\
    & = \frac{\alpha \delta^{k}}{(\delta - \alpha)^{k}} \Bigg[\frac{e^{-(\mu + \delta)t}}{\mu + \delta} \Bigg\{ \sum^{k-1}_{j=0} \Big( \frac{\delta - \alpha}{\mu + \delta} \Big)^j \sum^{j}_{i=0} \frac{t^i}{i!} (\mu + \delta)^{i} \Bigg\} - \frac{e^{-(\mu + \alpha)t}}{\mu + \alpha} \Bigg] +  \frac{\alpha}{\alpha + \mu} \Big( \frac{\delta}{\delta + \mu} \Big)^{k} \label{b_eq}
\end{align}
using Equations (\ref{a_eq}) and (\ref{c_eq}).\\

Now, suppose a bite at time $\tau_m \in (t_1, t_2]$ establishes precisely $Q=q$ hypnozoites in the liver. As per \textcite{mehra2021antibody}, since each hypnozoite activates in the interval $(\tau, t_2]$ independently with probability $B(t_2 - \tau)$, the number of hypnozoite activation events by time $t_2$, $M^r_m(t_2-\tau)$, has conditional distribution
\begin{align*}
    M_m(t_2-\tau) \sim \text{Binomial}(q, B(t_2 - \tau_m)),
\end{align*}
and thus conditional PGF
\begin{align}
    \EX \big[ z^{M_m(t_2-\tau)} | Q = q \big] = \big( 1 + B(t_2 - \tau_m) (z-1) \big)^q \label{m_q_cond_pgf}.
\end{align}
By the law of total expectation, noting that the size of the hypnozoite inoculum $Q$ is geometrically-distributed with mean $\nu$, it follows that $M_m(t_2 - \tau_m)$ has PGF
\begin{align}
    \EX \big[ z^{M_m(t_2-\tau_m)} \big] &= \sum^{\infty}_{q=0} \frac{1}{\nu + 1} \Big( \frac{\nu}{\nu + 1} \Big)^q   \EX \big[ z^{M_m(t_2-\tau)} | Q = q \big]  = \frac{1}{1 - \nu B(t_2 - \tau_m)(z-1) } \label{m_pgf}
\end{align}
where we have substituted Equation (\ref{m_q_cond_pgf}) and applied the geometric series summation. We note that Equation (\ref{m_pgf}) holds in the domain $|z| \leq 1$.\\

Assuming that dynamics of each relapse and primary infection are independent, $M_m(t_2 - \tau_m)$ and $V_m$ are independent random variables. Hence, the PGF for $(M_m(t_2 - \tau_m)+V_m)$, the total number of recurrences in the interval $[\tau_m, t_2)$ arising from a bite at time $\tau_m$, is given by the product
\begin{align}
    \EX \Big[z^{M_m(t_2 - \tau_m) + V_m} \Big] = \EX \Big[z^{M_m(t_2 - \tau_m)} \Big] \EX \Big[z^{V_m} \Big] = \frac{1 - p_\text{prim} + z p_\text{prim}}{1 - \nu B(t_2 - \tau_m)(z-1) } \label{bite_pgf_t1_t2}
\end{align}
where we have noted that $V_m \sim \text{Bernoulli}(p_\text{prim})$ and substituted Equation (\ref{m_pgf}).\\

Now, consider a mosquito bite that occurs at time $\tau_m \in (0, t_1]$, that is, prior to drug treatment. Any hypnozoites or infections arising from this bite may be affected by drug treatment. Here, we consider the random variable $(M^r(t_2 - \tau_m) - M^r(t_1 - \tau_m))$, which describes the number of relapses triggered by the bite in the interval $(t_1, t_2]$.\\

As before, we begin by examining the case of a single hypnozoite established in the liver at time $\tau_m \in (0, t_1]$, which will activate in the interval $(t_1, t_2)$ with probability
\begin{align*}
    p^r_A(t_2 - \tau_m, t_1 - \tau_m) &+ p^r_C(t_2 - \tau_m, t_1 - \tau_m) - p^r_A(t_1 - \tau_m, t_1 - \tau_m) + p^r_C(t_1 - \tau_m, t_1 - \tau_m)\\
    = & (1-p_\text{rad}) \big( B(t_2 - \tau) - B(t_1 - \tau) \big),
\end{align*}
where $B(t)$ is given by Equation (\ref{b_eq}).\\

Suppose that a bite at time $\tau_m \in (0, t_1]$ establishes $Q =q$ hypnozoites. Under the assumption that hypnozoite dynamics are i.i.d.,  the number of relapses $(M^r(t_2 - \tau_m) - M^r(t_1 - \tau_m))$ triggered in the interval $(t_1, t_2)$ has conditional distribution
\begin{align*}
    M^r_m(t_2 - \tau_m) - M^r_m(t_1 - \tau_m) \sim \text{Binomial} \big( q, (1-p_\text{rad})(B(t_2 - \tau_m) - B(t_1 - \tau_m)) \big).
\end{align*}

Using similar reasoning to Equations (\ref{m_q_cond_pgf}) and (\ref{m_pgf}) and applying the law of total expectation to account for a geometrically-distributed hypnozoite inoculum $Q$, it  follows that the random variable $M^r(t_2 - \tau_m) - M^r_m(t_1 - \tau_m)$ has PGF
\begin{align}
    \EX \big[ z^{M^r_m(t_2 - \tau_m) - M^r_m(t_1 - \tau_m)} \big] 
    = \frac{1}{1 - \nu (1-p_\text{rad}) (B(t_2 - \tau) - B(t_1 - \tau)) (z - 1) }, \label{bite_pgf_0_t1}
\end{align}
which likewise holds in the domain $|z| \leq 1$.\\

Equation (\ref{bite_pgf_0_t1}) characterises the number of recurrences arising from a single bite at time $\tau_m \in [0, t_1)$ initiated in the interval $(t_1, t_2)$, while Equation (\ref{m_pgf}) characterises the number of recurrences arising from a single bite at time $\tau_m \in [t_1, t_2)$ initiated in the interval $(\tau_m, t_2)$.\\ 

Next, we examine the distribution of bite times. Given $N(t_2) - N(t') = n$, from \textcite{lewis1967non}, the conditional distribution of bite times in the interval $[t', t_2)$ is equivalent to $n$ i.i.d. random variables with density
\begin{align*}
    f_1(\tau) = \frac{\lambda(\tau)}{m(t_2) - m(t')} \mathbbm{1}\{ \tau \in [t', t_2) \}.
\end{align*}

Hence, following a similar procedure to Section \ref{sec::baseline_inoc}, whereby we first condition on the number of bites in the interval, then the distribution of bite times, we have
\begin{align}
    \EX \big[ z^{I_{C_1}(t_1, t_2)} \big] & = \exp \bigg\{ -m(t_1) + \int^{t_1}_0 \lambda(\tau) \EX \Big[z^{M^r_m(t_2 - \tau) - M^r_m(t_1 - \tau)} \Big] d \tau \bigg\} \label{prelim_ic1} \\
    \EX \big[ z^{I_{C_2}(t_1, t_2)} \big] & = \exp \bigg\{ -(m(t_2) - m(t_1)) + \int^{t_2}_{t_1} \lambda(\tau) \EX \Big[z^{M_m(t_2 - \tau) + V_m} \Big] d \tau \bigg\} \label{prelim_ic2}.
\end{align}

From Equation (\ref{ic_prod}), since $I_{C_1}(t_1, t_2)$ and $I_{C_2}(t_1, t_2)$ are independent, it follows that
\begin{align}
    \EX \big[ z^{I_C(t_2) - I_C(t_1)} \big] = \exp \bigg\{ -m(t_2) + & \int^{t_1}_0 \lambda(\tau) \EX \Big[z^{M^r_m(t_2 - \tau) - M^r_m(t_1 - \tau)} \Big] d \tau +\\
    & \int^{t_2}_{t_1} \lambda(\tau) \EX \Big[z^{M_m(t_2 - \tau) + V_m} \Big] d \tau \bigg \}, \label{ir_pgf_prelim}
\end{align}
where we have substituted Equations (\ref{prelim_ic1}) and (\ref{prelim_ic2}).\\

Substituting Equations (\ref{bite_pgf_t1_t2}) and (\ref{bite_pgf_0_t1}) into Equation (\ref{ir_pgf_prelim}) yields the PGF of $I_C(t_2) - I_C(t_1)$:
\begin{align}
    \EX \big[ z^{I_C(t_2) - I_C(t_1)} \big] 
    =  \exp \bigg\{ -m(t_2) +  &\int^{t_1}_0 \frac{\lambda(\tau)}{1 - \nu (1 - p_\text{rad}) (B(t_2 - \tau) - B(t_1 - \tau)) (z - 1) } \, d \tau + \notag \\   
    & \int^{t_2}_{t_1} \frac{\lambda(\tau) (1-p_\text{prim} + z p_\text{prim})}{1 - \nu B(t_2 - \tau) (z - 1) } \, d \tau  \bigg\}, \label{ir_pgf}
\end{align}
which holds in the domain $|z| \leq 1$ and gives us the number of recurrences initiated in the interval $(t_1, t_2]$ following drug treatment at time $t_1$, where
\begin{itemize}
    \item $\lambda(\tau)$ is the infective mosquito bite rate, with the mean number of bites in the interval $(0, t]$ denoted by $m(t) = \int^t_0 \lambda(\tau) d \tau$;
    \item Hypnozoite inocula for each bite are geometrically distributed with mean $\nu$;
    \item Each bite triggers a primary infection with probability $p_\text{prim}$;
    \item $B(t')$ denotes the probability that a hypnozoite has activated time $t'$ after inoculation in the absence of treatment (Equation (\ref{b_eq})); and
    \item Each hypnozoite is killed instantaneously with probability $p_\text{rad}$ upon administration of radical cure at time $t_1$.
\end{itemize}

\section{Quantities of Epidemiological Significance} \label{sec::epi_q}

Using the PGFs derived in Section \ref{sec::inf_server_queue} (Equations (\ref{multi_pgf}), (\ref{multi_rad_pgf}) and (\ref{ir_pgf})), we can obtain several quantities of epidemiological significance pertaining to the relapse burden and the longer-term impacts of radical cure on the infection burden. Here, we consider an individual who is first exposed to infective mosquito bites at time zero, and is administered radical cure at time $t_1$.

\subsection{Size of Hypnozoite Reservoir} \label{subsec::num_hyp}

The risk of relapse for an individual is governed by the size of the hypnozoite reservoir. For short-latency strains, all hypnozoites in the liver (state $H$) may activate. For long-latency strains, in contrast, the hypnozoite reservoir is comprised of both dormant hypnozoites (states $1, \dots, k$) that can die, but are unable to activate; and non-latent hypnozoites (state $NL$) that have emerged from dormancy and may now activate. Here, we consider the total number of hypnozoites (state $H$) in the liver at time $t$, that is,
\begin{align*}
    N_H(t) = N_{NL}(t) + \sum^k_{m=1} N_m(t).
\end{align*}

The PGF for $N_H(t)$ follows from the joint PGFs for $\mathbf{N}(t)$ (denoted by $G$ in Equation (\ref{multi_pgf}), capturing hypnozoite dynamics in the absence of treatment) and $\mathbf{N}^{t_1}(t)$ (denoted by $G^{t_1}$ in Equation (\ref{multi_rad_pgf}), capturing dynamics following drug treatment at time $t_1$):
\begin{align}
    \EX \big[ z^{N_H(t)} \big] & = \begin{cases}
    G(t, z_1=z, \dots, z_k=z, z_{NL}=z, z_A=1, z_D=1, z_C=1, z_P=1, z_{PC}=1) & \text{ if } t < t_1\\
    G^{t_1}(t, z_1=z, \dots, z_k=z, z_{NL}=z, z_A=1, z_D=1, z_C=1, z_P=1, z_{PC}=1) & \text{ if } t \geq t_1
    \end{cases} \notag \\
    & =\begin{cases}
    \exp \Big\{ -m(t) + \int^t_0 \frac{\lambda(\tau)}{1 + \nu(1-z) p_H(t-\tau) } d \tau \Big\} & \text{ if } t < t_1\\
    \exp \Big\{ -m(t) + \int^{t_1}_0 \frac{\lambda(\tau)}{1 + \nu (1-p_\text{rad}) (1-z) p_H(t-\tau) } d \tau + \int^{t}_{t_1} \frac{\lambda(\tau)}{1 + \nu (1-z) p_H(t-\tau) } d \tau \Big\} & \text{ if } t \geq t_1
    \end{cases} \notag\\
    &= \exp \{ -m(t) + \ell(z, t)\} \label{H_pgf}
\end{align}
where
\begin{align*}
    \ell(z, t) = \begin{cases}
    \int^t_0 \frac{\lambda(\tau)}{1 + \nu(1-z) p_H(t-\tau) } d \tau & \text{ if } t < t_1\\
    \int^{t_1}_0 \frac{\lambda(\tau)}{1 + \nu (1-p_\text{rad}) (1-z) p_H(t-\tau) } d \tau + \int^{t}_{t_1} \frac{\lambda(\tau)}{1 + \nu (1-z) p_H(t-\tau) } d \tau &\text{ if } t \geq t_1.
    \end{cases}
\end{align*}

From the PGF given by Equation (\ref{H_pgf}), using Leibiniz integral rule, we can also compute the expected size of the hypnozoite reservoir
\begin{align}
        \EX[N_H(t)] &= \frac{\partial \EX [ z^{N_H(t)} ]}{\partial z} \bigg|_{z=1} \notag \\
        & = \begin{cases}
    \nu \int^t_0 \lambda(\tau) p_H(t-\tau) d \tau & \text{ if } t < t_1\\
    \nu (1-p_\text{rad}) \int^{t_1}_0 \lambda(\tau) p_H(t-\tau) d \tau + \nu \int^t_{t_1} \lambda(\tau) p_H(t-\tau) d \tau & \text{ if } t \geq t_1,\\
    \end{cases} \label{EX_H}
\end{align}
and the variance
\begin{align}
    \text{Var}(N_H(t)) &= \frac{\partial^2 \EX [ z^{N_H(t)} ]}{\partial z^2} \bigg|_{z=1} + \frac{\partial \EX [ z^{N_H(t)} ]}{\partial z} \bigg|_{z=1} - \Big( \frac{\partial \EX [ z^{N_H(t)} ]}{\partial z} \bigg|_{z=1} \Big)^2 \notag \\
    &= \begin{cases}
    \int^t_0 2 \lambda(\tau)  \big( \nu p_H(t-\tau) \big)^2 + \nu \lambda(\tau) p_H(t-\tau) d \tau & \text{ if } t < t_1\\
        \int^{t_1}_0 2 \lambda(\tau) \big( \nu (1-p_\text{rad})  p_H(t-\tau) \big)^2 + \nu (1-p_\text{rad}) \lambda(\tau) p_H(t-\tau) d \tau   & \text{ if } t \geq t_1 \\ +\int^t_{t_1} 2 \lambda(\tau) \big( \nu  p_H(t-\tau) \big)^2 +  \nu \lambda(\tau) p_H(t-\tau) d \tau. \\
    \end{cases} \label{Var_H}
\end{align}

To invert the PGF given by Equation (\ref{H_pgf}), we apply Fa\`{a} di Bruno's formula \parencite{di1857note}, allowing us to recover the PMF for $N_H(t)$ in terms of partial Bell polynomials $B^n_k$:
\begin{align}
    P \big( N_H(t) = n \big) &= \frac{e^{-m(t)}}{n!} \frac{d}{dz^n} \exp \big\{ \ell (z, t) \big\} \Big|_{z=0} \notag \\
    & = \frac{\exp \big\{ \ell(0, t) - m(t) \big\}}{n!} \sum^n_{k=1}  B_{n, k} \Big( \frac{\partial l}{\partial z}(0, t)), \frac{\partial^2 l}{\partial z^2}(0, t), \dots, \frac{\partial^{n-k+1} l}{\partial z^{n-k+1}}(0, t) \Big), \label{H_pmf}
\end{align}
where by Leibinz integral rule and the geometric summation, we have that
\begin{align}
    \frac{\partial^k l}{\partial z^k}(0, t) = \begin{cases}
     \nu^k k! \int^t_0  \frac{ \lambda(\tau) p_H(t-\tau)^k }{ [ 1+\nu p_H(t-\tau) ]^{k+1} } d \tau & \text{ if } t < t_1\\
    \nu^k k! \int^{t_1}_0  \frac{ \lambda(\tau) p_H(t-\tau)^k }{ [ 1+\nu p_H(t-\tau) ]^{k+1} } d \tau + \nu^k k! \int^t_{t_1}  \frac{ \lambda(\tau) (1-p_\text{rad})^k p_H(t-\tau)^k }{ [ 1+\nu (1-p_\text{rad}) p_H(t-\tau) ]^{k+1} } d \tau & \text{ if } t \geq t_1.
    \end{cases}
\end{align}

\subsubsection{Special Case: Short-Latency Hypnozoites, Constant Bite Rate}

In the simplest case, where we consider short-latency hypnozoites ($k=0$) and a constant bite rate $\lambda(t) = \lambda$ in the absence of radical cure, the PGF for the size of the hypnozoite reservoir at time $t$ can be evaluated (Equation (\ref{H_pgf}) in closed form
\begin{align}
    \EX \big[ z^{N_H(t)} \big] = \Big( \frac{1 + \nu (1-z) e^{-(\alpha + \mu) t}}{1 + \nu(1-z)} \Big)^\frac{\lambda}{\alpha + \mu}. \label{num_hyp_pgf_simple}
\end{align}

To compute the steady state distribution of $N_H^*$, we take the limit $t \to \infty$ in Equation (\ref{num_hyp_pgf_simple})
\begin{align}
    \EX \big[ z^{N_H^*} \big] = (1 + \nu - \nu z)^{-\frac{\lambda}{\alpha + \mu}}. \label{num_hyp_pgf_ss_simple}
\end{align}

Using the generalised binomial theorem, the PGF given by Equation (\ref{num_hyp_pgf_ss_simple}) can be inverted to yield the PMF for the size of the hypnozoite reservoir at steady state
\begin{align}
    P( N_H^* = n ) &= \frac{1}{n!} \frac{\nu^n}{(1 + \nu) ^{n+\frac{\lambda}{\alpha + \mu}}}  \Big( \frac{\lambda}{\alpha + \mu} + n - 1 \Big)_{(n)}
\end{align}
where $(x)_{(n)}$ denotes the Pochhammer symbol.

\subsubsection{Hypnozoite Reservoir Conditional on Infection Status}

We can also characterise the size of the hypnozoite reservoir conditional on the current infection status. Here, we revert to the general setting of a time-dependent bite rate $\lambda(t)$ and either short- or long-latency hypnozoites ($k \geq 0$). Suppose an individual does not have a blood-stage infection at time $t$, that is, $N_A(t) = N_P(t) =0$. By  \textcite{xekalaki1987method}, the conditional PGF for the size of the hypnozoite reservoir at time $t$, $N_H(t)$, can be obtained from the joint PGFs given by Equations (\ref{multi_pgf}) and (\ref{multi_rad_pgf}) as follows:
\begin{align}
    \EX \big[ z^{N_H(t)} | N_A(t) = N_P(t) = 0\big] &= 
    \begin{cases}
    \frac{G(z_1=z, \dots, z_k=z, z_{NL}=z, z_A=0, z_D=1, z_C=1, z_P=0, z_{PC}=1)}{G(z_1=1, \dots, z_k=1, z_{NL}=1, z_A=0, z_D=1, z_C=1, z_P=0, z_{PC}=1)} & \text{ if } t < t_1\\
    \frac{G^{t_1}(z_1=z, \dots, z_k=z, z_{NL}=z, z_A=0, z_D=1, z_C=1, z_P=0, z_{PC}=1)}{G^{t_1}(z_1=1, \dots, z_k=1, z_{NL}=1, z_A=0, z_D=1, z_C=1, z_P=0, z_{PC}=1)} & \text{ if } t \geq t_1
    \end{cases}\notag \\
    & = \exp \big\{  g(z, t) - g(1, t) \big\} \label{uninf_hyp_pgf}
\end{align}
where we define
\begin{align*}
    g(z, t) = \begin{cases}
    \int^t_0 \frac{\lambda(\tau) (1- p_\text{prim} e^{-\gamma (t-\tau)})}{1 + \nu p_H(t-\tau) (1-z) + \nu p_A(t-\tau) \big)} d \tau & \text{ if } t < t_1\\
    \int^t_{t_1} \frac{\lambda(\tau) (1- p_\text{prim} e^{-\gamma (t-\tau)})}{1 + \nu p_H(t-\tau) (1-z) + \nu p_A(t-\tau) \big)} d \tau + \int^{t_1}_0 \frac{\lambda(\tau) (1- p_\text{prim} ( 1 - p_\text{blood}) e^{-\gamma (t-\tau)})}{1 + \nu p^r_H(t-\tau, t_1 - \tau) (1-z) + \nu p^r_A(t-\tau, t_1 - \tau) \big)} d \tau & \text{ if } t \geq t_1
    \end{cases}.
\end{align*}

As before, the expected size of the hypnozoite reservoir in an uninfected individual can be obtained from Equation (\ref{uninf_hyp_pgf}) using Leibniz integral rule
\begin{align}
    \EX[ & N_H(t) | N_A(t) = N_P(t) = 0 ] = \frac{\partial \EX [ z^{N_H(t)} | N_A(t) = N_P(t) = 0]}{\partial z} \bigg|_{z=1} \notag \\
        & = \begin{cases}
    \nu \int^t_0 \frac{\lambda(\tau) p_H(t-\tau) (1- p_\text{prim} e^{-\gamma (t-\tau)}) }{[1 + \nu p_A(t-\tau)]^2} d \tau & \text{ if } t < t_1\\
    \nu (1-p_\text{rad}) \int^{t_1}_0 \frac{\lambda(\tau) p_H(t-\tau) (1- p_\text{prim} e^{-\gamma (t-\tau)}) }{[1 + \nu p^r_A(t-\tau, t_1 - \tau)]^2} d \tau + \nu \int^t_{t_1} \frac{\lambda(\tau) p_H(t-\tau) (1- p_\text{prim} e^{-\gamma (t-\tau)}) }{[1 + \nu p_A(t-\tau)]^2} d \tau & \text{ if } t \geq t_1.\\
    \end{cases} 
\end{align}

Similarly, we use Fa\`{a} di Bruno's formula \parencite{di1857note} to invert the conditional PGF given by Equation (\ref{uninf_hyp_pgf}) and obtain the conditional probability masses for $N_H(t)$, given an individual does not have an ongoing blood-stage infection at time $t$:
\begin{align}
    P \big( N_H(t) = n | & N_A(t) = N_P(t) = 0 \big) = \frac{1}{n!} \frac{\partial}{\partial z^n} \exp \big\{  g(z, t) - g(1, t) \big\} \Big|_{z=0} \notag \\
    & = \frac{\exp \big\{ g(0, t) - g(1, t) \big\}}{n!} \sum^n_{k=1} B_{n, k} \Big(  \frac{\partial g}{\partial z}(0,t), \frac{\partial^2 g}{\partial z^2}(0,t), \dots,  \frac{\partial^{n-k+1} g}{\partial z^{n-k+1}}(0, t) \Big) \label{uninf_hyp_pmf}
\end{align}
where, by Leibiniz integral rule and the geometric summation, we have that
\begin{align}
    \frac{\partial^k g}{\partial z^k} (0, t) &= \begin{cases}
    \nu^k k! \int^t_0  \frac{\lambda(\tau) (1- p_{\text{prim}} e^{-\gamma (t-\tau)})  p_H(t-\tau)^k }{ [ 1+\nu ( p_H(t-\tau) + p_A(t-\tau) ) ]^{k+1} }  d \tau & \text{ if } t < t_1\\
    \nu^k k! \big( \int^t_{t_1}  \frac{\lambda(\tau) (1- p_{\text{prim}} e^{-\gamma (t-\tau)})  p_H(t-\tau)^k }{ [ 1+\nu ( p_H(t-\tau) + p_A(t-\tau) ) ]^{k+1} }  d \tau  & \text{ if } t \geq t_1\\
    \qquad  + \int^t_{t_1}  \frac{\lambda(\tau) (1- p_{\text{prim}} (1-p_\text{blood}) e^{-\gamma (t-\tau)})  p^r_H(t-\tau, t_1 - \tau)^k }{ [ 1+\nu ( p^r_H(t-\tau, t_1 - \tau) + p^r_A(t-\tau, t_1 - \tau) ) ]^{k+1} }  d \tau \big)  
    \end{cases}\label{uninf_hyp_pmf_int}
\end{align}

The conditional PMF for the size of the hypnozoite reservoir at time $t$, $N_H(t)$, given an ongoing blood stage infection at time $t$, that is, $N_A(t) + N_P(t) > 0$, follows readily from Equations (\ref{uninf_hyp_pmf}) and (\ref{no_A_P}) (see Section \ref{subsec::relapse_contrib})
\begin{align}
    P(N_H(t) = n | & N_A(t) + N_P(t) > 0) \notag\\
    = & \frac{P(N_H(t) = n) - P(N_H(t) =n | N_A(t) = N_P(t) =0) P(N_A(t) = N_P(t) = 0)}{1 - P(N_A(t) = N_P(t) = 0)}
\end{align}

\subsection{Probability of Infection} \label{subsec::relapse_contrib}

Prior to the establishment of a hypnozoite reservoir, primary infections are likely to be the dominant source of infection for an individual in an endemic setting. As the hypnozoite reservoir accrues over time, we expect relapses to contribute to an increasingly large proportion of the infection burden; however, we expect primary infections to once again become the dominant source of infection if the hypnozoite reservoir is substantially reduced due to radical cure. We thus begin by examining the probability of relapse and primary infection both before and after the administration of drug treatment.\\

From the joint PGFs for $\mathbf{N}(t)$, (denoted $G$ in Equation (\ref{multi_pgf})), which holds prior to drug treatment, and $\mathbf{N}^{t_1}(t)$, (denoted $G^{t_1}$ in Equation (\ref{multi_rad_pgf})), which holds after drug treatment at time $t_1$, the probability that an individual does not have an ongoing primary infection at time $t$ is
\begin{align}
    P( & N_P(t) = 0) = \EX \big[ z^{N_P(t) } \big] \Big|_{z=0} \notag \\ 
    &= \begin{cases}
    G(t, z_1=1, \dots, z_k=1, z_{NL}=1, z_A=1, z_D=1, z_C=1, z_P=0, z_{PC}=1) & \text{ if } t < t_1\\
    G^{t_1}(t, z_1=1, \dots, z_k=1, z_{NL}=1, z_A=1, z_D=1, z_C=1, z_P=0, z_{PC}=1) & \text{ if } t \geq t_1
    \end{cases}  \notag\\
    &= \begin{cases}
    \exp \Big\{- p_\text{prim} \int^t_0 \lambda(\tau) e^{-\gamma(t-\tau)} d \tau \Big\} & \text{ if } t < t_1\\
    \exp \Big\{- p_\text{prim} \Big[ \int^t_{t_1} \lambda(\tau) e^{-\gamma(t-\tau)} d \tau + (1-p_\text{blood}) \int^{t_1}_0 \lambda(\tau) e^{-\gamma(t-\tau)} \Big] \Big\} & \text{ if } t \geq t_1.
    \end{cases}  \label{no_P}
\end{align}

Similarly, the probability that the individual does not have an ongoing relapse at time $t$ is
\begin{align}
    P( & N_A(t)  = 0) = \EX \big[ z^{N_A(t)} \big] \Big|_{z=0} \notag \\ 
    &= \begin{cases}
    G(t, z_1=1, \dots, z_k=1, z_{NL}=1,, z_A=0, z_D=1, z_C=1, z_P=1, z_{PC}=1) &\text{ if } t < t_1\\
    G^{t_1}(t, z_1=1, \dots, z_k=1, z_{NL}=1, z_A=0, z_D=1, z_C=1, z_P=1, z_{PC}=1) &\text{ if } t \geq t_1
    \end{cases}
     \notag\\
    & = \begin{cases}
    \exp \Big\{ -m(t) + \int^t_0 \frac{\lambda(\tau)}{1 + \nu p_A(t-\tau)} d \tau \Big\} &\text{ if } t < t_1\\
    \exp \Big\{ -m(t) + \int^t_{t_1} \frac{\lambda(\tau)}{1 + \nu p_A(t-\tau) } d \tau + \int^{t_1}_0 \frac{\lambda(\tau)}{1 + \nu p^r_A (t-\tau, t_1-\tau) } d \tau \Big\} &\text{ if } t \geq t_1.
    \end{cases} \label{no_A}
\end{align}

The probability that the individual is neither experiencing a relapse, nor a primary infection at time $t$ is given by
\begin{align}
    P(&N_A(t) = 0, N_P(t) = 0) = \EX \big[ z^{N_P(t) + N_A(t)} \big] \Big|_{z=0} \notag \\ 
    & = \begin{cases}
     G(t, z_1=1, \dots, z_k=1, z_{NL}=1, z_A=0, z_D=1, z_C=1, z_P=0, z_{PC}=1) &\text{ if } t < t_1\\
    G^{t_1}(t, z_1=1, \dots, z_k=1, z_{NL}=1, z_A=0, z_D=1, z_C=1, z_P=0, z_{PC}=1) &\text{ if } t \geq t_1
    \end{cases}
     \notag\\
    & = \begin{cases}
    \exp \Big\{ -m(t) + \int^t_0 \frac{\lambda(\tau)(1- p_\text{prim} e^{-\gamma(t-\tau)} )}{1 + \nu p_A(t-\tau) } d \tau \Big\} &\text{ if } t < t_1\\
    \exp \Big\{ -m(t) + \int^t_{t_1} \frac{\lambda(\tau)(1 - p_\text{prim} e^{-\gamma(t-\tau)})}{1 + \nu p_A(t-\tau) } d \tau + \int^{t_1}_0 \frac{\lambda(\tau) (1-p_\text{prim} (1-p_\text{blood}) e^{-\gamma(t-\tau)} )}{1 + \nu p^r_A(t-\tau, t_1-\tau) } d \tau \Big\} &\text{ if } t \geq t_1.
    \end{cases} \label{no_A_P}
\end{align}

Equations (\ref{no_P}) to (\ref{no_A_P}) allow us to obtain several parameters describing the relative contributions of relapses and primary infections to blood-stage infection, noting that hypnozoite activation events in close proximity to mosquito bites can result in a multiple infections consisting of overlapping relapses and primary infections. Suppose an individual has an an ongoing blood-stage infection at time $t$, that is, $\{ N_A(t) + N_P(t)>0 \}$. We can compute the probability that the blood-stage infection is due to
\begin{itemize}
    \item Hypnozoite activation (i.e. relapse) only: 
    \begin{align}
        P(N_P(t)=0, N_A(t)>0 | N_P(t) + N_A(t) > 0) = \frac{P(N_P(t)=0) - P(N_P(t) = N_A(t) = 0)}{1-P(N_P(t) = N_A(t) = 0)} \label{prob_relapse_only}
    \end{align}
    \item Reinfection (i.e. primary infection) only:
    \begin{align}
        P(N_P(t)>0, N_A(t)=0 | N_P(t) + N_A(t) > 0) = \frac{P(N_A(t)=0) - P(N_P(t) = N_A(t) = 0)}{1-P(N_P(t) = N_A(t) = 0)} \label{prob_prim_only}
    \end{align}
    \item Hypnozoite activation and reinfection (i.e. overlapping relapse and primary infection):
    \begin{align}
        P(N_P(t)>0, N_A(t)>0 &| N_P(t) + N_A(t) > 0) \notag \\
        & = \frac{1 - P(N_P(t)=0) - P(N_A(t)=0) + P(N_P(t) = N_A(t) = 0)}{1-P(N_P(t) = N_A(t) = 0)} \label{prob_prim_relapse}
    \end{align}
\end{itemize}

\subsection{Multiple Infections} \label{subsec::moi}
Hypnozoite activation or reinfection events in quick succession may give rise to overlapping blood-stage infections, which can involve the co-circulation of genetically-distinct parasite strains in the bloodstream \parencite{fola2017higher}. Here, we assume that each primary infection is comprised of a single parasite strain, while hypnozoites are all genetically heterologous. The multiplicity of infection  (MOI) is therefore given by the total number of active infections (i.e. primary infections and relapses) at time $t$: $M_I(t) := N_A(t) + N_P(t)$. Using the joint PGFs for $\mathbf{N}(t)$ (denoted $G$ in Equation (\ref{multi_pgf})), which holds prior to drug treatment and $\mathbf{N}^{t_1}(t)$ (denoted $G^{t_1}$ in Equation (\ref{multi_rad_pgf})), which holds following drug treatment at time $t_1$, the PGF for $M_I(t)$ is given by
\begin{align}
    \EX[z^{M_I(t)}] & = \begin{cases}
    G(t, z_1=1, \dots, z_k=1, z_{NL}=1, z_A=z, z_D=1, z_C=1, z_P=z, z_{PC}=1) & \text{ if } t < t_1\\
    G^{t_1}(t, z_1=1, \dots, z_k=1, z_{NL}=1, z_A=z, z_D=1, z_C=1, z_P=z, z_{PC}=1) & \text{ if } t \geq t_1
    \end{cases} \notag \\
    & = \begin{cases}
    \exp \Big\{ -m(t) + \int^t_0 \lambda(\tau) \frac{  z p_\text{prim} e^{-\gamma (t-\tau)} + (1- p_\text{prim} e^{-\gamma (t-\tau)})}{1 + \nu p_A(t-\tau) (1-z) } d \tau \Big\} & \text{ if } t < t_1 \\
     \exp \Big\{ -m(t) + \int^t_{t_1} \lambda(\tau) \frac{z p_\text{prim} e^{-\gamma (t-\tau)} + (1- p_\text{prim} e^{-\gamma (t-\tau)})}{1 + \nu p_A(t-\tau) (1-z) } d \tau & \text{ if } t \geq t_1 \\
      \qquad \qquad + \int^{t_1}_{0} \lambda(\tau) \frac{z p_\text{prim} ( 1 - p_\text{blood}) e^{-\gamma (t-\tau)} + (1- p_\text{prim} ( 1 - p_\text{blood}) e^{-\gamma (t-\tau)}) }{1 + \nu p_A^r(t-\tau, t_1-\tau) (1-z) } d \tau \Big\} 
    \end{cases} \label{moi_pgf} \\
    &= \exp\{ -m(t) + f(z, t) \}, \notag 
\end{align}
where we denote
\begin{align*}
    f(z, t) &= \begin{cases}
    \int^t_0 \lambda(\tau) \frac{z p_\text{prim} e^{-\gamma (t-\tau)} + (1- p_\text{prim} e^{-\gamma (t-\tau)})}{1 + \nu p_A(t-\tau) (1-z) } d \tau & \text{ if } t < t_1 \\
    \int^t_{t_1} \lambda(\tau) \frac{z p_\text{prim} e^{-\gamma (t-\tau)} + (1- p_\text{prim} e^{-\gamma (t-\tau)})}{1 + \nu p_A(t-\tau) (1-z) } d \tau & \text{ if } t \geq t_1 \\
     \qquad +  \int^{t_1}_{0} \lambda(\tau) \frac{z p_\text{prim} ( 1 - p_\text{blood}) e^{-\gamma (t-\tau)} +  (1- p_\text{prim} ( 1 - p_\text{blood}) e^{-\gamma (t-\tau)})}{1 + \nu p_A^r(t-\tau, t_1-\tau) (1-z) } d \tau
    \end{cases}
\end{align*}

To recover the probability mass function for $M_I(t)$ from the PGF given in Equation (\ref{moi_pgf}), we use a similar procedure to that in Section \ref{subsec::num_hyp}. By Fa\`{a} di Bruno's formula \parencite{di1857note}, we have that
\begin{align}
    P \big( M_I(t) = n \big) &= \frac{e^{-m(t)}}{n!} \frac{d^n}{dz^n} \exp \big\{ f(z, t) \big\} \Big|_{z=0} \notag \\
    & = \frac{\exp \big\{ f(0, t) - m(t) \big\} }{n!} \sum^n_{k=1} B_{n, k} \bigg( \frac{\partial f}{\partial z}(0, t), \frac{\partial^2 f}{\partial z^2}(0, t), \dots, \frac{\partial f^{(n-k+1)}}{\partial z^{(n-k+1)}}(0, t) \bigg) \label{moi_pmf}
\end{align}
where $B^n_k$ denote the partial Bell polynomials and by Leibinz integral rule,
\begin{align}
    \frac{\partial^k f}{\partial z^k}(0, t) = \begin{cases}
    k! \int^t_0 \frac{\lambda(\tau) [\nu p_A (t-\tau) ]^{k-1} }{ [ 1+\nu p_A(t-\tau)]^{k} } \Big( p_\text{prim} e^{-\gamma (t-\tau)} + \frac{\nu p_A(t-\tau) (1- p_\text{prim} e^{-\gamma (t-\tau)})}{1 + \nu p_A(t-\tau)} \Big) d \tau & \text{ if } t < t_1\\
    k! \Big( \int^t_{t_1} \frac{\lambda(\tau) [\nu p_A (t-\tau) ]^{k-1} }{ [ 1+\nu p_A(t-\tau)]^{k} } \Big( p_\text{prim} e^{-\gamma (t-\tau)} + \frac{\nu p_A(t-\tau) (1- p_\text{prim} e^{-\gamma (t-\tau)})}{1 + \nu p_A(t-\tau)} \Big) d \tau  & \text{ if } t \geq t_1 \\ 
    \hspace{1.5mm} +  \int^{t_1}_{0} \frac{\lambda(\tau) [\nu p_A (t-\tau) ]^{k-1} }{ [ 1+\nu p_A(t-\tau)]^{k} } \Big( p_\text{prim} (1 - p_\text{blood}) e^{-\gamma (t-\tau)} + \frac{\nu p_A(t-\tau) (1- p_\text{prim} (1 - p_\text{blood}) e^{-\gamma (t-\tau)})}{1 + \nu p_A(t-\tau)} \Big) d \tau \Big)
    \end{cases}
\end{align}

\subsection{Time to First Recurrence} \label{subsec::time_first_recur}

Next, we consider the time to first recurrence following drug treatment, a quantity that has been surveyed in longitudinal epidemiological studies across a range of transmission settings \parencite{robinson2015strategies, taylor2019resolving, corder2020quantifying}. Randomised controlled trials, comparing the time to first recurrence following treatment with radical cure compared to blood-stage treatment only, have also been used to quantify the efficacy of radical cure \parencite{nelwan2015randomized}.\\

Evaluating the PGF for the number of recurrences following drug treatment (Equation (\ref{ir_pgf})) at $z=0$ yields the probability of no recurrences in the interval $[t_1, t_2)$, given drug treatment is administered at time $t_1$:
\begin{align}
    P \big( & I_C(t_2) - I_C(t_1) = 0 \big) = \EX \big[ z^{I_C(t_2)-I_C(t_1)} \big] \Big|_{z=0} \notag \\ 
    &= \exp \bigg\{ -m(t_2) +  \int^{t_1}_0 \frac{\lambda(\tau)}{1 + \nu (1-p_\text{rad}) (B(t_2 - \tau) - B(t_1 - \tau)) } \, d \tau + \int^{t_2}_{t_1} \frac{\lambda(\tau) (1-p_\text{prim})}{1 + \nu B(t_2 - \tau) } \, d \tau  \bigg\}. \label{eq:first_recur_gen}
\end{align}

Since recurrences include both primary infections and relapses, setting $p_\text{prim}=0$ in Equation (\ref{eq:first_recur_gen}) yields the distribution for the time to first relapse following drug treatment.   

\subsection{Cumulative Number of Infections Over Time} \label{subsec::cum_inf}

To quantify the longer-term impacts of a single administration of radical cure, we seek to compare the infection burden following radical cure, as opposed to blood-stage treatment only. Here, we consider the cumulative number of infections experienced in the interval $(t_1, t_2]$, following drug treatment at time $t_1$, $I_C(t_2)-I_C(t_1)$. The PGF for $I_C(t_2)-I_C(t_1)$ (Equation (\ref{ir_pgf})) can be written
\begin{align*}
    \EX \big[ z^{I_C(t_2)-I_C(t_1)} \big] = \exp\{ -m(t_2) + h(z, t_1, t_2) \}
\end{align*}
where
\begin{align*}
    h(z, t_1, t_2) = \int^{t_1}_0 \frac{\lambda(\tau)}{1+\nu (1-p_\text{rad}) [ B(t_2 - \tau) - B(t_1 - \tau)](1-z)} d \tau + \int^{t_2}_{t_1} \frac{\lambda(\tau)[ 1 - p_\text{prim} + z p_\text{prim}]}{1 + \nu B(t_2-\tau)(1-z)} d \tau.
\end{align*}

We can thus compute the expected number of infections following drug treatment
\begin{align}
    \EX[ & I_C(t_2) - I_C(t_1)] = \frac{\partial \EX [ z^{I_C(t_2) - I_C(t_1)} ]}{\partial z} \bigg|_{z=1} \notag \\
    & = 
    \nu (1-p_\text{rad}) \int^{t_1}_0 \lambda(\tau) \big( B(t_2 - \tau) - B(t_1 - \tau) \big) d \tau + \int^{t_2}_{t_1} \lambda(\tau) \big( p_\text{prim} + \nu B(t_2-\tau) \big) d \tau, \label{EX_IC}
\end{align}
as well as the variance
\begin{align}
    \text{Var} & (I_C(t_2) - I_C(t_1)) = \frac{\partial^2 \EX [ z^{I_C(t_2) - I_C(t_1)} ]}{\partial z^2} \bigg|_{z=1} + \frac{\partial \EX [ z^{I_C(t_2) - I_C(t_1)} ]}{\partial z} \bigg|_{z=1} - \Big( \frac{\partial \EX [ z^{I_C(t_2) - I_C(t_1)} ]}{\partial z} \bigg|_{z=1} \Big)^2 \notag \\
    =& \int^{t_1}_0 \lambda(\tau) \Big[ 2 \big( \nu (1-p_\text{rad})  (B(t_2 - \tau) - B(t_1 - \tau)) \big)^2 + \nu (1-p_\text{rad})  (B(t_2 - \tau) - B(t_1 - \tau)) \Big] d \tau \notag \\
    & \vspace{3mm} + \int^{t_2}_{t_1} \lambda(\tau) \Big[ 2  \big( \nu B(t_2-\tau) \big)^2 + (2 p_\text{prim} + 1) \nu B(t_2-\tau) + p_\text{prim} \Big] d \tau. \label{Var_IC}
\end{align}

As in Section \ref{subsec::num_hyp}, we can invert the PGF in Equation (\ref{ir_pgf}) to yield the PMF for $I_C(t_2) - I_C(t_1)$ in terms of partial Bell polynomials $B^n_k$ by applying Fa\`{a} di Bruno's formula \parencite{di1857note}:
\begin{align}
    P \big( & I_C(t_2) - I_C(t_1) = n \big) = \frac{e^{-m(t_2)}}{n!} \frac{d^n}{dz^n} \exp \big\{ h(z, t_1, t_2) \big\} \Big|_{z=0} \notag \\
    & = \frac{\exp \big\{ h(0, t_1, t_2) - m(t_2)) \big\} }{n!} \sum^n_{k=1} B_{n, k} \bigg( \frac{\partial h}{\partial z}(0, t_1, t_2), \frac{\partial^2 h}{\partial z^2}(0, t_1, t_2), \dots, \frac{\partial h^{(n-k+1)}}{\partial z^{(n-k+1)}}(0, t_1, t_2) \bigg) \label{cum_inf_pmf}
\end{align}
where, using Leibiniz integral rule and the geometric series summation
\begin{align}
    \frac{\partial^k h}{\partial z^k} = &k! \int^{t_2}_{t_1} \frac{\lambda(\tau) [v B (t_2-\tau) ]^{k-1} }{ [ 1+\nu B(t_2-\tau)]^{k} } \Big( p_\text{prim} + \frac{\nu B(t_2-\tau) (1- p_\text{prim})}{1 + \nu B(t_2-\tau)} \Big) d \tau + \notag \\
    & k! \int^{t_1}_0 \frac{\lambda(\tau) [\nu (1-p_\text{rad}) ( B(t_2-\tau) - B(t_1 - \tau) ) ]^{k} }{ [ 1+\nu (1-p_\text{rad}) ( B(t_2-\tau) - B(t_1 - \tau) )]^{k+1} } d \tau.
\end{align}

\section{Illustrative Results} \label{sec::illustrative_results}

In Section \ref{sec::epi_q}, we derived several quantities of epidemiological significance pertaining to the dynamics of the hypnozoite reservoir and the infection burden in a general transmission setting. Comparing these dynamics following radical cure, as opposed to blood-stage treatment only, can help elucidate the epidemiological effects of radical cure. Here, we provide illustrative results for hypnozoite and infection dynamics for both short-latency (tropical) and long-latency (temperate) strains. For simplicity, we restrict our attention to a constant transmission setting.\\

\subsection{Hypnozoite Distributions are Zero-Inflated at Early Times and in Low Transmission Settings}

\begin{figure}
    \centering
    \includegraphics[width=\textwidth]{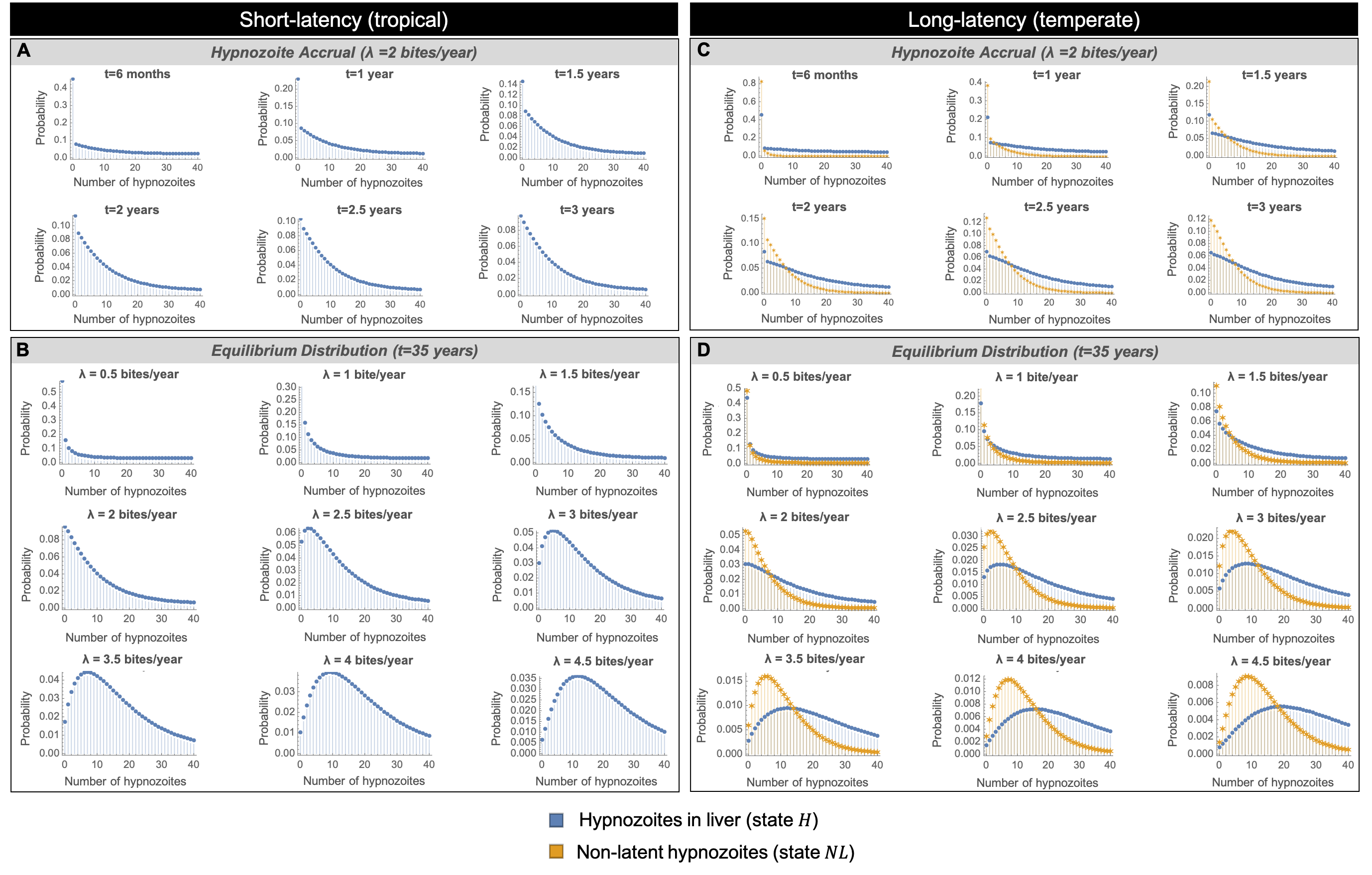}
    \caption{Distributions for the size of the hypnozoite reservoir in the absence of drug treatment (Equation (\ref{H_pmf})). The total number of hypnozoites in the liver (both latent and non-latent) are shown in blue, while non-latent hypnozoites are shown in orange. For long-latency strains, the hypnozoite reservoir is comprised of both latent hypnozoites (that may die, but not activate) and non-latent hypnozoites (that may either activate or die); for short-latency strains, all hypnozoites in the liver are subject to both death and activation (Section \ref{sec::baseline_hyp}). In subplots A and C, we show the accrual of the hypnozoite reservoir over time ($t=0.5,1,1.5,2,2,5,3$ years) in a constant transmission setting with bite rate $\lambda=2/365$ day$^{-1}$. In subplots B and D, we show the equilibrium distribution ($t=35$ years) for the size of the hypnozoite reservoir for various biting intensities ($\lambda=0.5,1,1.5,2,2,5,3,3.5,4,4.5$ bites/year). Based on estimates from \textcite{white2014modelling}, we assume an average of $\nu=9$  hypnozoites are established per bite. Hypnozoite activation and clearance rates, $\alpha=1/334$ day$^{-1}$, $\mu=1/442$ day$^{-1}$, and, in the case of long-latency strains, the number of latency compartments $k=35$, as well as the rate of progression through successive hypnozoite strains $\delta=1/5$ day$^{-1}$, have been taken from \textcite{white2014modelling}.}
    \label{fig:num_hyp_illustrative}
\end{figure}

We begin by considering the size of the hypnozoite reservoir in the absence of drug treatment (Figure \ref{fig:num_hyp_illustrative}). Distributions for the total size of the hypnozoite reservoir (state $H$) are shown in blue; non-latent hypnozoites (state $NL$), which govern the risk of relapse but account for only a subset of the reservoir for long-latency strains, are shown in orange. The progressive accrual of the hypnozoite reservoir in an intermediate transmission setting is demonstrated in Figures \ref{fig:num_hyp_illustrative}A and \ref{fig:num_hyp_illustrative}C. PMFs for the size of the hypnozoite reservoir are distinctly zero-inflated at early times, since, while each bite establishes a sizeable hypnozoite inoculum on average, there is a reasonably high probability of an individual having experienced no mosquito bites early on. Due to the enforced dormancy period, there is a delay of approximately six months before a non-latent hypnozoite reservoir starts to accumulate (Figure \ref{fig:num_hyp_illustrative}C).  However, the hypnozoite reservoir, both latent and non-latent, eventually stabilises in size as the clearance of hypnozoites from the liver (through either death or activation) offsets replenishment of the reservoir through mosquito inoculation. In Figures \ref{fig:num_hyp_illustrative}B and \ref{fig:num_hyp_illustrative}D, we examine the equilibrium hypnozoite distribution ($t=35$ years) across a range of biting intensities. For very low bite rates, we likewise observe zero-inflated distributions since the probability of a recent mosquito bite (relative to the expected duration of hypnozoite carriage) is comparatively low. Non-zero central tendencies emerge as the biting intensity increases and the number of recent bites contributing to the hypnozoite reservoir is expected to increase.

\subsection{Multiple Infections are Driven By Relapses}

\begin{figure}
    \centering
    \includegraphics[width=\textwidth]{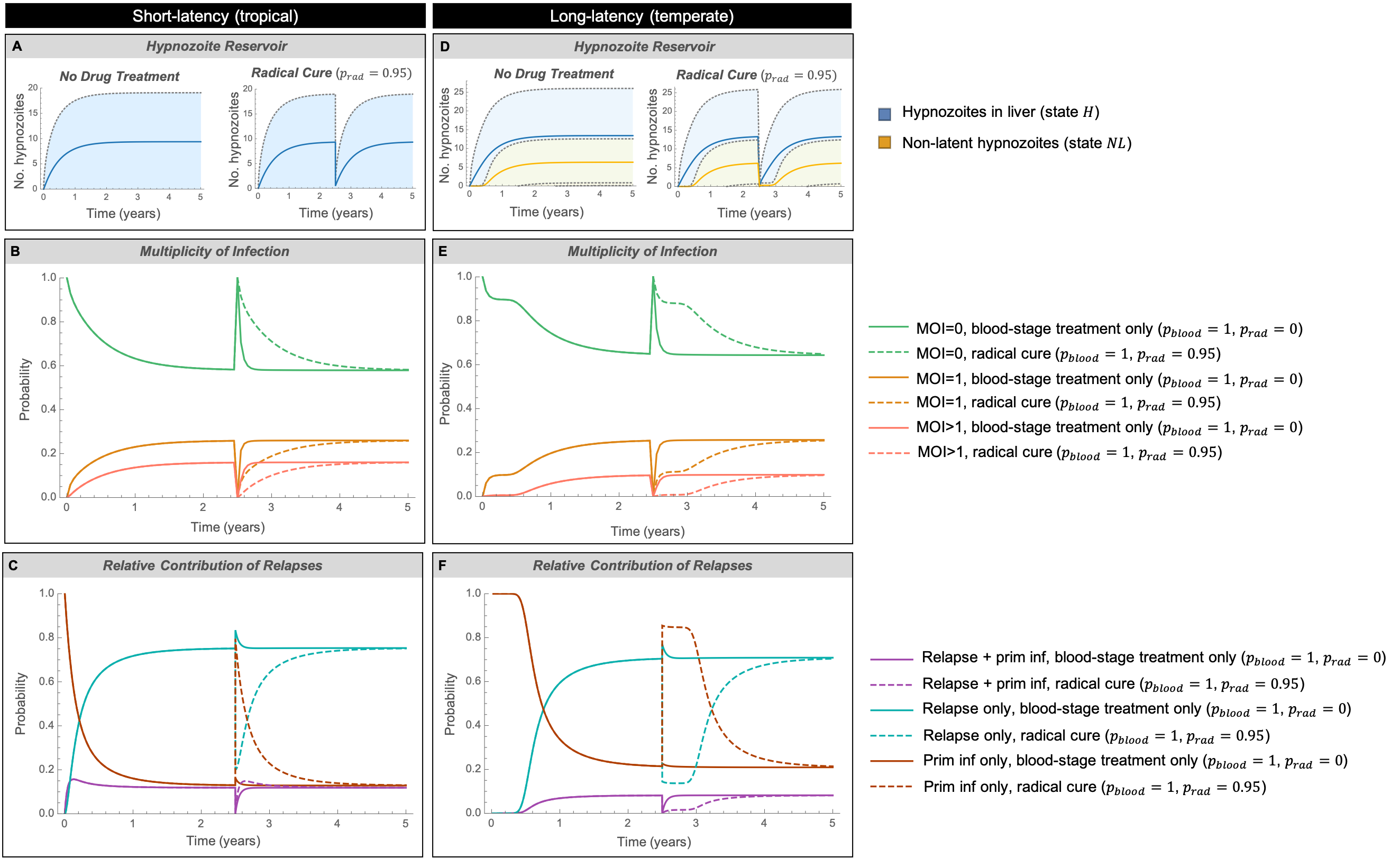}
    \caption{Hypnozoite and infection dynamics for an individual in a general transmission setting. At time zero, we assume an individual first enters an endemic setting. At time $t=2.5$ years, we consider the administration of either radical cure (dotted lines, $p_\text{rad}=0.95$,  $p_\text{blood}=1$) or blood-stage treatment only (solid lines, $p_\text{rad}=0$, $p_\text{blood}=1$). In subplots A and D, we show the expected size of the hypnozoite reservoir (Equation (\ref{EX_H})), with shaded regions indicating one standard deviation above and below the mean (Equation (\ref{Var_H})). Distributions for the multiplicity of infection, that is, the number of active infections $N_A(t)+N_P(t)$ over time (Equation (\ref{moi_pmf})) are shown in subplots B and E. Given an ongoing infection (that is, $N_A(t)+N_P(t)>0$), the probability of that infection comprising of a relapse only, primary infection only or overlapping relapses and primary infections (Equations (\ref{prob_prim_only}), (\ref{prob_relapse_only}) and (\ref{prob_prim_relapse}), computed using Equations (\ref{no_P}), (\ref{no_A}) and (\ref{no_A_P})) is shown in subplots C and F. Here, we assume a constant transmission setting with bite rate $\lambda=2/365$ day$^{-1}$ and all bites necessarily leading to a primary infection, that is, $p_\text{prim}=1$. At baseline, we assume that each primary infection and relapse is cleared at rate $\gamma=1/20$ day$^{-1}$. Based on estimates from \textcite{white2014modelling}, we assume an average of $\nu=9$  hypnozoites are established per bite. Hypnozoite activation and clearance rates, $\alpha=1/334$ day$^{-1}$, $\mu=1/442$ day$^{-1}$, and, in the case of long-latency strains, the number of latency compartments $k=35$, as well as the rate of progression through successive hypnozoite strains $\delta=1/5$ day$^{-1}$, are from \textcite{white2014modelling}.}
    \label{fig:moi_illustrative}
\end{figure}

Multiple (overlapping) infections can arise from reinfection and hypnozoite activation events in quick succession, with experimental data revealing polyclonal relapses even in the absence of reinfection \parencite{popovici2018genomic}. Here, we characterise the relationship between the size of the hypnozoite reservoir; the prevalence of multiple infections, and the relative contribution of relapses to the infection burden. Figures \ref{fig:moi_illustrative}A and \ref{fig:moi_illustrative}D illustrate the expected size of the hypnozoite reservoir over time, with hypnozoite distributions expected to stabilise prior to the administration of drug treatment at $t_1=2.5$ years. In Figures \ref{fig:moi_illustrative}B and \ref{fig:moi_illustrative}E, we examine distributions for the multiplicity of infection (MOI) in a single individual, under the assumption that each primary infection is comprised of a single clone, while all hypnozoites are genetically heterologous (our definition of MOI is equivalent to the number of active infections at a given point in time under our model). The conditional probability of an active infection comprising of either a primary infection (brown), relapse (turquoise) or both (purple) is shown in Figures \ref{fig:moi_illustrative}C and \ref{fig:moi_illustrative}F.\\

Primary infections are initially the dominant source of infection (Figures \ref{fig:moi_illustrative}C and \ref{fig:moi_illustrative}F, solid brown line), as the hypnozoite reservoir is yet to accumulate (Figures \ref{fig:moi_illustrative}A and \ref{fig:moi_illustrative}D). There is thus a low initial probability of multiple infections (MOI$>$1) since overlapping primary infections, arising from mosquito bites in quick succession, are unlikely to occur under the given bite rate (Figures \ref{fig:moi_illustrative}B and \ref{fig:moi_illustrative}E, solid red line).  Given long-latency hypnozoites necessarily undergo a dormancy phase before they may activate, there is a delay of approximately six months before there is a non-negligible risk of relapse, during which we expect only primary infections (Figure \ref{fig:moi_illustrative}F, solid brown line) with MOI=1 (Figure \ref{fig:moi_illustrative}E, solid orange line). As the hypnozoite reservoir accrues and eventually stabilises in size (Figure \ref{fig:moi_illustrative}A, Figure \ref{fig:moi_illustrative}D), both the risk of multiple infections (MOI$>$1) (Figures \ref{fig:moi_illustrative}B and \ref{fig:moi_illustrative}E, solid red line) and the relative contributions of relapses to the infection burden (Figures \ref{fig:moi_illustrative}C and \ref{fig:moi_illustrative}F, solid turquoise line) rise steadily, before plateauing.\\

Upon the administration of drug treatment at time $t_1=2.5$ years after the individual first enters the endemic setting, all ongoing recurrences are instantaneously cleared. Since blood-stage treatment does not clear the hypnozoite reservoir, the probability of infection increases sharply to its original level within weeks of treatment (Figures \ref{fig:moi_illustrative}B and \ref{fig:moi_illustrative}E, solid lines), with relapses remaining the dominant source of infection (Figures \ref{fig:moi_illustrative}C and \ref{fig:moi_illustrative}F, solid turquoise line). Following treatment with radical cure, however, there is period during which primary infections dominate (Figures \ref{fig:moi_illustrative}C and \ref{fig:moi_illustrative}F, dotted brown line), which is longer for long-latency strains since hypnozoites must emerge from dormancy prior to activation; the risk of infection prior to drug treatment is only reached after a year has passed, when the hypnozoite reservoir is expected to have been replenished to its previous level (Figures \ref{fig:moi_illustrative}A and \ref{fig:moi_illustrative}D).\\

\subsection{Heterogeneity in Recurrences Following Drug Treatment}

\begin{figure}
    \centering
    \includegraphics[width=\textwidth]{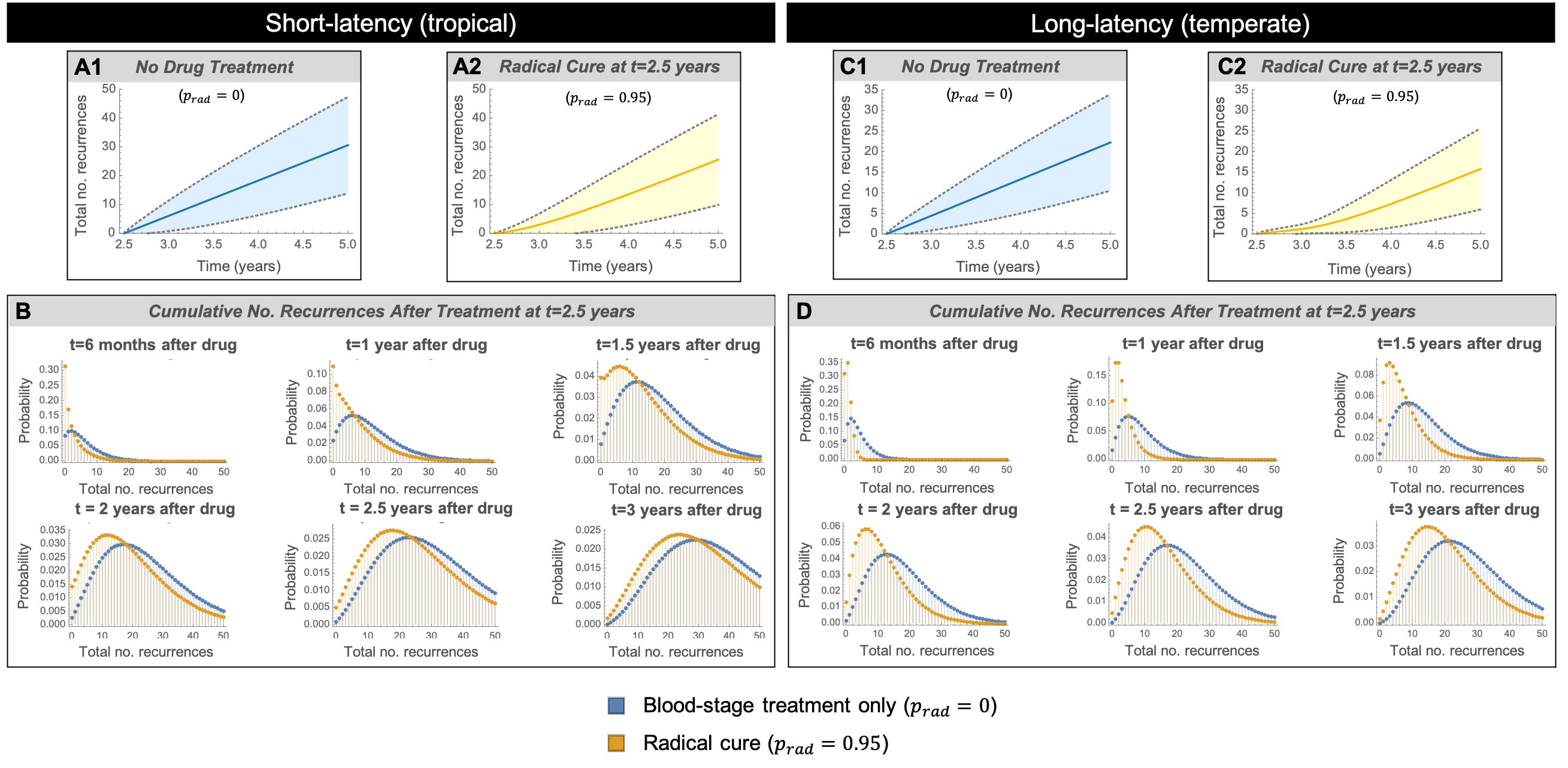}
    \caption{Distributions for the cumulative number of recurrences following drug treatment. In subplots A and C, we compare a scenario with no drug treatment (A1, C1) against the administration of reasonably efficacious radical cure ($p_\text{rad}=0.95$) (A2, C2) at time $t_1=2.5$ years after an individual has been first exposed to an endemic setting; expected values (Equation (\ref{EX_IC})), with shaded regions indicating one standard deviation (Equation (\ref{Var_IC})) above and below the mean, are shown. For long-latency strains, a single administration of radical cure in this simulation is expected to prevent $6.4$ relapses over a $2.5$ year period following treatment, while  for short-latency strains, we expect $5$ relapses to be prevented. PMFs for the cumulative number of infections in the interval $(t_1, t_1+t]$ following drug treatment (Equation (\ref{cum_inf_pmf})) are shown in subplots B and D for various time points ($t=0.5,1,2,5,2,2.5,3$ years). Here, we assume a constant transmission setting with bite rate $\lambda=2/365$ day$^{-1}$ and all bites necessarily leading to a primary infection, that is, $p_\text{prim}=1$. Model parameters ($\gamma$, $\mu$, $\alpha$, $k$, $\delta$) as per Figure \ref{fig:moi_illustrative}.}
    \label{fig:cumul_inf_illustrative}
\end{figure}

To quantify the longer-term effects of a single administration of radical cure on the infection burden, we examine the cumulative number of recurrences following drug treatment (Figure \ref{fig:cumul_inf_illustrative}). Since we do not account for a drug washout period, blood-stage treatment only ($p_\text{rad}=0$) clears ongoing recurrences upon administration, but does not affect subsequent infections. Prior to drug treatment at $t_1=2.5$ years, we expect both the size of the hypnozoite reservoir (Figures \ref{fig:moi_illustrative}A and \ref{fig:moi_illustrative}D) and the probability of infection (Figures \ref{fig:moi_illustrative}B and \ref{fig:moi_illustrative}E, solid lines) to have largely stabilised. Accordingly, the expected number of recurrences following blood-stage treatment only at time $t_1=2.5$ years is approximately linear (Figures \ref{fig:cumul_inf_illustrative}A1 and \ref{fig:cumul_inf_illustrative}C1). However, distributions for the cumulative number of relapses following blood-stage treatment only (Figures \ref{fig:cumul_inf_illustrative}B and \ref{fig:cumul_inf_illustrative}D, blue curves) reveal substantial heterogeneity, arising from the batch arrival of hypnozoites for each bite; while infective bites are relatively infrequent, each bite establishes a reasonably large hypnozoite inoculum and thus contributes substantially to the relapse burden.\\

Following treatment with radical cure, there is a delay before the hypnozoite reservoir is replenished, and consequently a period during which relapses are limited (Figures \ref{fig:moi_illustrative}C and \ref{fig:moi_illustrative}F, dotted lines). Hence, the expected number of recurrences following radical cure (Figures \ref{fig:cumul_inf_illustrative}A2, \ref{fig:cumul_inf_illustrative}C2) initially rises slowly due to primary infections dominating, but similarly becomes approximately linear as the hypnozoite reservoir accrues and stabilises in size (Figures \ref{fig:moi_illustrative}A and \ref{fig:moi_illustrative}D). For long-latency strains, the expected number of relapses grows slowly for a prolonged period (Figure \ref{fig:cumul_inf_illustrative}C2) as each hypnozoite established in the liver must emerge from dormancy prior to contributing to the risk of relapse.

\section{Discussion} \label{sec::discussion}

The hypnozoite reservoir governs the epidemiology of \textit{P. vivax}, with important implications for treatment and control. Radical curative therapies, which target the hypnozoite reservoir, have the potential to aid elimination efforts. Here, we have developed a stochastic within-host model to capture hypnozoite and infection dynamics for vivax malaria in a general transmission setting, whilst accounting for the administration of radical cure. We have proposed a relapse-clearance model adapted to both short- and long-latency hypnozoite strains, that extends previous models \parencite{white2014modelling, mehra2020activation} to allow for drug treatment and an exponentially-distributed relapse following each hypnozoite activation event. Extending our previous work \parencite{mehra2021antibody} to concurrently monitor hypnozoite and infection dynamics, we have embedded our relapse-clearance model in an epidemiological framework capturing repeated mosquito inoculation. By constructing an open network of infinite server queues with batch arrivals, we have derived joint PGFs for the size of the hypnozoite reservoir and the cumulative number of infections over time, both in the absence of drug treatment (Equation (\ref{multi_pgf})) and following the administration of radical cure (Equation (\ref{multi_rad_pgf})), yielding analytic distributions for several quantities of epidemiological significance.\\

Although the risk of relapse is dependent on the dynamics of the hypnozoite reservoir, a common approach across many statistical and transmission models has been an assumed distributional form for the risk of relapse \parencite{ishikawa2003mathematical, aguas2012modeling, roy2013potential, chamchod2013modeling, lover2014distribution, robinson2015strategies, white2016variation, taylor2019resolving}. Efforts to explicitly model the accrual of the hypnozoite reservoir, with clearance (through either death or activation) offsetting replenishment (through mosquito bites), have been more limited. By embedding an activation-clearance model for short-latency strains in a population-level transmission model allowing for variable hypnozoite inocula per bite, \textcite{white2014modelling} have obtained distributions for prevalence and the size of the hypnozoite reservoir under a range of control measures, including radical cure. Here, we jointly characterise within-host hypnozoite and infection dynamics for both short- and long-latency strains, whilst accounting for the effects of radical cure. To our knowledge, we provide the first analytical descriptions of several important epidemiological quantities, including the size of the hypnozoite reservoir; distributions of multiple infections; the relative contributions of primary infections to the infection burden and the cumulative number of infections over time. By describing the time evolution of the hypnozoite reservoir in a general transmission setting, we capture transient dynamics that are unlikely to be captured by an assumed distributional form, but provide insight into the epidemiological consequences of radical cure. Our model can be calibrated efficiently to data using the time to first recurrence following drug treatment, a frequently collected piece of epidemiological information \parencite{robinson2015strategies, taylor2019resolving, corder2020quantifying} for which we provide explicit analytic formulae (Equation (\ref{eq:first_recur_gen})). Our model thus has the potential to address questions around the heterogeneity of relapse risk in communities. While our within-host model provides insight into the epidemiological effects of radical cure on a single individual, population-level models are required to evaluate the utility of radical cure as a tool for elimination and control \parencite{robinson2015strategies, white2018mathematical}. Our analytic within-host distributions can be readily embedded in multiscale models to yield further insights.\\

Our model is underpinned by various simplifying assumptions. Similarly to \textcite{white2014modelling} and \textcite{mehra2020activation}, our relapse-clearance model for a single hypnozoite considers a baseline scenario, with spontaneous hypnozoite activation assumed to occur at a constant rate post-dormancy; we do not consider other possible mechanisms that have been hypothesised to temporarily elevate reactivation rates, such as systemtic febrile illness \parencite{shanks2013activation} or bites from certain mosquito vectors \parencite{hulden2011activation}. We model drug treatment by assuming a fixed probability of survival for each hypnozoite, and a fixed probability of persistence for each blood-stage infection, under the assumption that the effects of drug treatment are instantaneous. Drug washout periods, with antimalarial half-lives ranging from approximately 40 minutes for artensunate \parencite{morris2011review}, to 6 hours for primaquine (radical cure) \parencite{white1992antimalarial} and 30-60 days for chloroquine \parencite{white1992antimalarial}, can be incorporated into our relapse-clearance framework using an appropriate forcing function. In fact, the form of the instantaneous forcing function in Equations (\ref{rad_deq_L1}) to (\ref{rad_deq_D}) follows from taking the limit
\begin{align*}
    \lim_{\Delta t \to 0} \eta(\Delta t) H(t - s_j) H(s_j + \Delta t - t) = \ln \big( (1 - p)^{-1} ) \delta_D(t - s_j),
\end{align*}
where $H$ denotes the Heavyside step function (not to be confused with the hypnozoite state $H$); $\delta_D( \cdot )$ denotes the Dirac delta function (not to be confused with $\delta$, a scalar parameter that denotes the rate of transition between successive latency compartments), and the rate constant $\eta(\Delta t)$ is chosen such that a drug causes a transition with probability $p$ in the interval $(s_j, s_j + \Delta t)$
\begin{align*}
    e^{-\eta(\Delta t) \Delta t} = 1 - p \iff \eta(\Delta t) = \frac{\ln \big( (1-p)^{-1} \big)}{\Delta t}.
\end{align*}
While drug washout periods are an important consideration for interpreting data for the time to first recurrence, in this work, we have been concerned primarily with longer-term dynamics following drug treatment.\\

Although we account for multiple infections arising from reinfection (mosquito bites) or hypnozoite activation events in quick succession, our model does not capture the complexities of blood-stage infection. Under our framework, an infection refers to a period of parasitemia triggered by either by the activation of a single hypnozoite (relapse), or a single mosquito bite (primary infection); we model neither parasite densities, nor clinical disease status, over the course of each infection. The assumption that the duration of each infection is exponentially-distributed can be relaxed, but independent clearance of each infection is critical to our analytic framework; as such, our model is not equipped to capture phenomena like within-host competition between co-circulating strains \parencite{roode2005dynamics}. Our model, moreover, does not account for the acquisition of immunity. While antimalarial immunity can modulate parasite clearance \parencite{artavanis2003war}, we assume that the duration of each infection is identically-distributed. Given the progressive acquisition of immunity against clinical disease (clinical immunity), followed by the modulation of parasitemia (anti-parasite immunity) \parencite{schofield2006clinical}, modelling immunity requires the coupling of our framework to mechanistic models of blood-stage infection. Since antimalarial immunity is strain-specific, albeit with cross-protectivity amongst strains \parencite{mueller2013natural}, strain structure is another important consideration; repeated exposure to a single strain, either through the activation of homologous hypnozoites or across multiple bites, can generate strong strain-specific immune protection, but partial protection against heterologous strains \parencite{mueller2013natural}, thereby modulating blood-stage dynamics. Modelling strain structure would also allow for precise distributions of MOI (here, we assume MOI to be given by the number of active infections at a given point in time, without accounting for the possibility of overlapping strains across infections, nor polyclonal primary infections).\\

Our work, nonetheless, characterises the dependence of infection dynamics for \textit{P. vivax}, including distributions of multiple infections, the relative contribution of relapses to the infection burden and the cumulative number of infections over time, on the accrual of the hypnozoite reservoir. By comparing the infection burden under in the absence of drug treatment against the administration of arbitrarily effective radical cure, our work contributes to the epidemiological understanding of the effects of radical cure on \textit{P. vivax} malaria.

\section*{Acknowledgements}

D. Khoury's research is supported by the Australian Research Council (ARC) (DP180103875) and the National Health and Medical Research Council (NHMRC) of Australia (1141921). J.M. McCaw's research is supported by the ARC (DP170103076).  J.A. Flegg's research is supported by the ARC (DE160100227, DP200100747).

\newpage

\printbibliography

\end{document}